\documentclass[prd,aps,nofootinbib]{revtex4}
\usepackage{amssymb,amsmath,epsfig}
\usepackage[bookmarks]{hyperref}
\usepackage[usenames,dvipsnames]{color}

\newcommand{\Real}{\mathbb{R}}

\newcommand{\diag}{\mbox{diag}}
\newcommand{\divrg}{\mbox{div}\,}

\newcommand{\proof}{\noindent {\bf Proof. }}

\newcommand{\qed}{\hfill \fbox{} \vspace{.3cm}}

\newtheorem{lemma}{Lemma}
\newtheorem{proposition}{Proposition}

\newtheorem{theorem}{Theorem}

\begin{document}

\title{The geometry of the tangent bundle and the relativistic kinetic theory of gases}

\author{Olivier Sarbach and Thomas Zannias}

\affiliation{Instituto de F\'\i sica y Matem\'aticas,
Universidad Michoacana de San Nicol\'as de Hidalgo,\\
Edificio C-3, Ciudad Universitaria, 58040 Morelia, Michoac\'an, M\'exico.}

\begin{abstract}
This article discusses the relativistic kinetic theory for a simple collisionless gas from a geometric perspective. We start by reviewing the rich geometrical structure of the tangent bundle $TM$ of a given spacetime manifold, including the splitting of the tangent spaces of $TM$ into horizontal and vertical subspaces and the natural metric and symplectic structure it induces on $TM$. Based on these structures we introduce the Liouville vector field $L$ and a suitable Hamiltonian function $H$ on $TM$. The Liouville vector field turns out to be the Hamiltonian vector field associated to $H$. On the other hand, $H$ also defines the mass shells as Lorentzian submanifolds of the tangent bundle. When restricted to these mass shells, the projection of the integral curves of $L$ on the base manifold describes a family of future directed timelike geodesics. A simple collisionless gas is described by a distribution function on a particular mass shell, satisfying the Liouville equation. Together with the Liouville vector field the distribution function can be thought of as defining a fictitious incompressible fluid on the mass shells, with associated conserved current density. Flux integrals of this current density provide the averaged properties of the gas, while suitable fibre integrals of the distribution function define divergence-free tensor fields on the spacetime manifold such as the current density and stress-energy tensor. Finally, we discuss the relationship between symmetries of the spacetime manifold and symmetries of the distribution function. Taking advantage of the natural metric and symplectic structure on $TM$, we show that groups of isometries $G$ of the spacetime manifold lift naturally to groups of isometries and symplectic flows on the tangent bundle. Motivated by these properties, we define a distribution function to be $G$-invariant whenever it is invariant under the lifted isometries.

As a first application of our formalism we derive the most general spherically symmetric distribution function on any spherically symmetric spacetime and write the Einstein-Liouville equations as effective field equations on the two-dimensional radial manifold. As a second application we derive the most general collisionless distribution function on a Kerr black hole spacetime background.
\end{abstract}

\date{\today}


\maketitle
\section{Introduction}

The covariant kinetic theory of relativistic gases has been founded in a fundamental paper by Synge written in 1934~\cite{jS34}. In this paper, Synge introduces the world lines of the gas particles as the fundamental ingredient for the development of the theory. Through this idea emerges the concept of the invariant distribution function and a statistical description of a relativistic gas. However, at that epoch, no one would have thought that such a theory would find any applications in the physical universe, and thus Synge's idea had been left largely unexplored.

The situation changed dramatically after the 1960s. New technologies led to more accurate observations which in turn resulted in the discovery of quasars, radio pulsars, rapidly fluctuating X-ray sources and of course the cosmic microwave background radiation. These new discoveries led to the revival of Synge's idea and to the modern development of relativistic kinetic theory. The formulation of kinetic theory based on the relativistic Boltzmann equation has been initiated as early as $1961$, see the paper by Tauber and Weinberg~\cite{gTjW61}. Work by Israel~\cite{wI63} derived conservation laws based on the fully covariant Boltzmann equation and the relativistic version of the H-theorem. Moreover, the important notion of the state of thermodynamical equilibrium in a gravitational field has been clarified, and it has been recognized in~\cite{wI63} that a perfect gas has a bulk viscosity, a purely relativistic effect. Relativistic kinetic theory supported the formulation of transient thermodynamics~\cite{wI76,wIjS76,wIjS79a,wIjS79b,wHlL83,wHlL85}. The particular case of photon gases and its properties has been analyzed by Lindquist~\cite{rL66}. At the theoretical forefront, the development of black hole physics and the realization that the interaction of black holes with the rest of the universe requires a fully general relativistic treatment gave further impetus for the development of relativistic kinetic theory. Nowadays it has evolved into an important branch of relativistic astrophysics and cosmology. For an overview, see the books by Cercignani and Kremer~\cite{CercignaniKremer-Book} and Ringstr\"om~\cite{Ringstrom-Book}.

Relativistic kinetic theory not only finds applications in modeling astrophysical and cosmological scenarios, but also leads to new challenges in the field of mathematical relativity. The formulation of the Cosmic Censorship Hypothesis motivated the search of matter models that go beyond the traditional fluids or magneto-fluids, and here studies of the Einstein-Liouville, Einstein-Maxwell-Vlasov and Einstein-Boltzmann equations are relevant and on the frontier of studies in mathematical relativity, see for instance Refs.~\cite{dByC73,aR04,gRaR92,nNmT09} and Ref.~\cite{hA11} for a recent review. Kinetic theory offers a good candidate for a matter model that avoids pathologies such as shocks and shell-crossing singularities. Studies of solutions of the Einstein-Liouville system are characterized by a number of encouraging properties, see for instance~\cite{mDaR07}, \cite{hA11} and \cite{Ringstrom-Book}.

Motivated by the above considerations, in recent work~\cite{oStZ13}, we gave a mathematically oriented introduction to the relativistic kinetic theory of a simple gas based on Synge's ideas and inspired by early work by Ehlers~\cite{jE71,jE73}. In the approach of~\cite{oStZ13} (see also chapter 3 in Ref.~\cite{cL11}) the Poincar\'e one-form of the tangent bundle $TM$ led naturally to the symplectic form on $TM$, which in combination with an appropriate Hamiltonian function $H$, gave rise to the Liouville vector field $L$ identified as the corresponding Hamiltonian vector field on $TM$. When $L$ is restricted to particular energy surfaces of constant $H$, called the mass shells, the projections of the integral curves of $L$ on the spacetime manifold $M$ define a family of future directed timelike geodesics. In this framework, a relativistic simple gas, that is, a collection of neutral, spinless classical particles of the same positive rest mass $m > 0$, is described by a distribution function on the associated mass shell $\Gamma_m$. 

The analysis in~\cite{oStZ13} has been based solely on the symplectic structure of $TM$ induced by the Poincar\'e one-form. As it turned out, the symplectic structure was sufficient to introduce a volume form on $TM$, which in conjunction with the Hamiltonian, also induced a volume form on the mass shell. The presence of the volume form on the mass shell is crucial for the physical interpretation of the distribution function. As discussed in~\cite{oStZ13} the distribution function combined with that framework leads to the construction of a number of observables such as the particle current density and stress-energy tensor. Whenever collisions are present and the gas is dilute enough so that the molecular chaos hypothesis is reasonable, then the distribution function satisfies the Boltzmann equation. For the case where the self-gravity of the gas is important the Einstein-Boltzmann equations need to be considered. For the particular case of a self-gravitating collisionless simple gas, the collision integral can be neglected and the system reduces to the Einstein-Liouville system of equations.

In this article, we provide additional insights into the relativistic kinetic theory which complements the symplectic formulation in~\cite{oStZ13} and reinforces the relation between the rich geometric structure of the tangent bundle and the structure of kinetic theory. In particular, we show that the entire theory can be deduced from the splitting of the tangent space of $TM$ into horizontal and vertical subspaces induced by the Levi-Civita connection of the spacetime manifold $(M,g)$. An almost complex structure can be introduced on $TM$ which rotates horizontal into vertical vectors and vice versa. Moreover, the splitting leads to the presence of a natural metric $\hat{g}$ on the tangent bundle $TM$. In fact, this metric $\hat{g}$ has been introduced a long time ago by Sasaki~\cite{sS58} in the context of Riemannian manifolds, and its connection to relativistic kinetic theory has already been pointed out by Lindquist~\cite{rL66}. In the present work, we show that the splitting of $TM$ and the bundle metric $\hat{g}$ play a key role in the description and physical interpretation of the relativistic kinetic theory of gases.

This is accomplished by first showing that $\hat{g}$ together with the almost complex structure give rise to a symplectic form. This symplectic form turns out to be equivalent to the one obtained from the Poincar\'e one-form employed in our previous approach~\cite{oStZ13}. Next, the Liouville vector field $L$ is introduced, and shown to generate geodesics on $(TM,\hat{g})$. In a next step, we utilize the metric $\hat{g}$ and the Liouville vector field to define the Hamiltonian function $H:=\hat{g}(L,L)/2$ on $TM$ whose level surfaces define preferred submanifolds of $TM$. Particular families of these surfaces define the mass shell $\Gamma_m$ as Lorentzian submanifolds of $TM$. In particular, there is a natural associated volume form on $\Gamma_m$ which allows one to define integrals of functions on the mass shell.

The central assumption of relativistic kinetic theory is that the averaged properties of the gas are described by a one-particle distribution function $f$ on the mass shell $\Gamma_m$. In order to interpret the distribution function within our approach, we point out that $f$ together with the Liouville vector field $L$ describe a fictitious incompressible fluid on $\Gamma_m$ with associated current density ${\cal J} = f L/m$. Due to the presence of the induced metric on $\Gamma_m$, this current density ${\cal J}$ combined with a suitable spacelike hypersurface $\Sigma$ in $\Gamma_m$ gives rise to a flux integral. This flux integral is physically interpreted as providing a new definition of the averaged number of occupied trajectories that intersect $\Sigma$. For a collisionless gas this implies immediately that ${\cal J}$ is divergence-free, which in turn implies that the distribution function $f$ must obey the Liouville equation $\pounds_L f = 0$. Furthermore, the current density ${\cal J}$ on $\Gamma_m$ gives rise to a physical current density $J$ on the spacetime manifold $(M,g)$ which is also conserved. This physical current represents the first moment of the distribution function through a fibre integral. Higher moments of the distribution function can be constructed in an analogous way and shown to be conserved as well. Of particular relevance is the second moment which gives rise to the stress-energy tensor, and allows one to couple gravity to the kinetic gas through Einstein's field equations.

The insights gained from our geometric formulation of kinetic theory are helpful in various aspects. As we show, they lead to a clear understanding of the relationship between the symmetries of the spacetime manifold $(M,g)$ and those of the distribution function. In particular, we show that groups of isometries of the spacetime manifold give rise to groups of isometries of $(TM,\hat{g})$ and to symplectic flows on $TM$. As an application of these properties we derive the most general spherically symmetric distribution function on an arbitrary spherically symmetric spacetime manifold and also the most general solution of the Liouville equation on a Kerr black hole background. These new results should have a wide range of applications in astrophysically interesting scenarios and will be explored in future work.

This article is structured as follows: in the next section we introduce the tangent bundle and discuss some of its properties that are relevant to this article. In particular, we discuss the connection map and the way that this map leads to a natural splitting of the tangent space of $TM$ into horizontal and vertical subspaces and to the almost complex structure $J$ on $TM$. In Section~\ref{Sec:SSMetric} we introduce the metric $\hat{g}$ on the tangent bundle and discuss its most important properties, including the fact that together with $J$ it gives rise to a symplectic form on $TM$. Next, in Section~\ref{Sec:KTheory} we discuss the basic ingredients required for our geometric formulation of the relativistic kinetic theory of a simple gas, culminating in our new interpretation of the distribution function and the derivation of the Liouville equation for a collisionless gas. In the same section, we also discuss the construction of observables and comment on the Newtonian limit of the Einstein-Liouville equations.

Next, in Section~\ref{Sec:Symmetries} we discuss the important connection between groups of isometries of the background spacetime $(M,g)$ and symmetries of the distribution function. In particular, we discuss the structure of distribution functions which share the symmetries of the background spacetime. In Section~\ref{Sec:Applications} we discuss applications of our formalism. As a first concrete example we discuss the structure of the most general spherically symmetric distribution function on an arbitrary spherically symmetric spacetime and reformulate the Einstein-Liouville system of equations as an effective problem on the two-dimensional radial manifold. As a second example we derive the most general collisionless distribution function on a Kerr black hole background. Conclusions and an outlook towards future work are given in Section~\ref{Sec:Conclusions}. A technical proof is included in the appendix.

For generality purposes, we develop the theory on a spacetime $(M,g)$ of an arbitrary dimension $n\geq 2$. We employ the following notation and conventions: $(M,g)$ denotes a $C^\infty$-differentiable, $n$-dimensional Lorentzian manifold with the signature convention $(-,+,+,\ldots,+)$ for the metric. We use the Einstein summation convention with Greek indices $\mu,\nu,\sigma,\ldots$ running from $0$ to $d:=n-1$ and Latin indices $i,j,k,\ldots$ running from $1$ to $d$. On any $C^\infty$- differentiable manifold $N$, ${\cal F}(N)$ and ${\cal X}(N)$ denote the class of $C^\infty$-differentiable functions and vector fields, respectively. The operators $i_X$ and $\pounds_X$ refer to the interior product and Lie derivative,  respectively, with respect to the vector field $X$. Round brackets enclosing indices refer to total symmetrization, for example $v_{(ij)} := (v_{ij} + v_{ji})/2$. We use units for which $c=1$.

\section{On the tangent bundle of a Lorentzian manifold}

In this section, we review some basic properties of the tangent bundle $TM$ associated to an arbitrary, $n$-dimensional spacetime manifold $(M,g)$. In particular, we introduce the connection map which will play an important role in what follows, and we discuss the corresponding splitting of the tangent space at any point of $TM$ into a horizontal and a vertical subspace. This splitting is induced by the connection map in conjunction with the push-forward of the natural projection.

\subsection{Basic properties and adapted local coordinates}

Let $T_x M$ denote the vector space of all tangent vectors $p$ at some event $x\in M$. The tangent bundle of $M$ is defined as
\begin{displaymath}
TM := \{ (x,p) : x\in M, p\in T_x M \},
\end{displaymath}
with the associated projection map $\pi: TM\to M$, $(x,p)\mapsto x$. The fibre at $x\in M$ is the space $\pi^{-1}(x) = (x,T_x M)$ which is naturally  isomorphic to $T_x M$. The first basic property of the tangent bundle is described in the following lemma. Although it constitutes a standard result (see, for instance, Ref.~\cite{DoCarmo-Book1}) we include the proof for completeness of the presentation.

\begin{lemma}
\label{Lem:TM}
$TM$ is an orientable, $2n$-dimensional $C^\infty$-differentiable manifold.
\end{lemma}

\proof The proof is based on the observation that a local chart $(U,\phi)$ of $M$ defines in a natural way a local chart $(V,\psi)$ of $TM$ as follows. Let $V:=\pi^{-1}(U)$ and define
\begin{eqnarray*}
\psi: V &\to& \phi(U)\times \Real^n\subset \Real^{2n},\\
(x,p) &\mapsto& \left( x^0,\ldots,x^d,p^0,\ldots,p^d \right)
:= \left( \phi(x), dx_x^0(p),\ldots,dx_x^d(p) \right)
\end{eqnarray*}
with inverse
\begin{eqnarray*}
\psi^{-1}: \phi(U)\times \Real^n &\to& V,\\
(x^0,\ldots,x^d,p^0,\ldots,p^d) &\mapsto& 
\left( \phi^{-1}(x^0,\ldots,x^d), 
 p^\sigma\left. \frac{\partial}{\partial x^\sigma} \right|_{\phi^{-1}(x^0,\ldots,x^d)} \right).
\end{eqnarray*}
By taking an atlas $(U_\alpha,\phi_\alpha)$ of $M$, the corresponding local charts $(V_\alpha,\psi_\alpha)$ cover $TM$.

Next, suppose $V_{\alpha\beta} := V_\alpha\cap V_\beta\neq \emptyset$, and consider the corresponding transition map $\psi_{\alpha\beta} := \psi_\beta\circ\psi_\alpha^{-1} : \psi_\alpha(V_{\alpha\beta})\subset\Real^{2n} \to \psi_\beta(V_{\alpha\beta})\subset\Real^{2n}$. We show that this map is $C^\infty$-differentiable and that its Jacobian matrix has positive determinant, implying that the collection $(V_\alpha,\psi_\alpha)$ constitutes an oriented atlas of $TM$. Denoting by $(x^\mu)$ and $(y^\mu)$ the local coordinates of $M$ associated to $(U_\alpha,\phi_\alpha)$ and $(U_\beta,\phi_\beta)$, we find
\begin{displaymath}
\psi_{\alpha\beta}(x^\nu,p^\nu) = 
 \left( y^0(x^\nu),\ldots,y^d(x^\nu),
 p^\sigma \left. \frac{\partial y^0}{\partial x^\sigma} \right|_{\phi_\alpha^{-1}(x^\nu)},
 \ldots,
 p^\sigma \left. \frac{\partial y^d}{\partial x^\sigma} \right|_{\phi_\alpha^{-1}(x^\nu)} \right),
\end{displaymath}
where $y^\mu(x^\nu) := \phi_\beta\circ\phi_\alpha^{-1}(x^\nu)$ is the transition function on $M$ and $\left( \left. \frac{\partial y^\mu}{\partial x^\nu} \right|_x \right)$ is the corresponding Jacobian matrix. Since these are $C^\infty$-functions by the differentiability of $M$, it follows that $\psi_{\alpha\beta}$ is also $C^\infty$-differentiable. Finally, the Jacobian determinant of $\psi_{\alpha\beta}$ is
\begin{displaymath}
\det\left( D\psi_{\alpha\beta}(x^\nu,p^\nu) \right)
 = \det\left( \begin{array}{ll}
\left. \frac{\partial y^\mu}{\partial x^\nu} \right|_{\phi_\alpha^{-1}(x^\nu)} & 0 \\
p^\sigma\left. \frac{\partial^2 y^\mu}{\partial x^\sigma\partial x^\nu} 
\right|_{\phi_\alpha^{-1}(x^\nu)} & 
\left. \frac{\partial y^\mu}{\partial x^\nu} \right|_{\phi_\alpha^{-1}(x^\nu)}
\end{array} \right)
 = \left[ \det\left( \left. \frac{\partial y^\mu}{\partial x^\nu} \right|_{\phi_\alpha^{-1}(x^\nu)}
 \right) \right]^2,
\end{displaymath}
which is positive. Notice that this proof implies that $TM$ is always orientable, regardless of whether or not $M$ is orientable.
\qed

We call $(x^\mu,p^\mu)$ \emph{adapted local coordinates}. $\left\{ \left. \frac{\partial}{\partial x^\mu} \right|_{(x,p)},\left. \frac{\partial}{\partial p^\mu} \right|_{(x,p)} \right\}$ and\\ $\left\{ dx^\mu_{(x,p)},dp^\mu_{(x,p)} \right\}$ are the corresponding basis of the tangent and cotangent spaces of $TM$ at $(x,p)$. Any tangent vector $Z\in T_{(x,p)}(TM)$ can then be expanded as
\begin{displaymath}
Z = X^\mu\left. \frac{\partial}{\partial x^\mu} \right|_{(x,p)}
 + Y^\mu\left. \frac{\partial}{\partial p^\mu} \right|_{(x,p)},\quad
X^\mu = dx^\mu_{(x,p)}(Z),\quad Y^\mu = dp^\mu_{(x,p)}(Z).
\end{displaymath}
Likewise, any covector $\omega\in T^*_{(x,p)}(TM)$ can be expanded as
\begin{displaymath}
\omega = \alpha_\mu\left. dx^\mu \right|_{(x,p)}
 + \beta_\mu\left. dp^\mu \right|_{(x,p)},\quad
 \alpha_\mu = \omega\left( \left. \frac{\partial}{\partial x^\mu} \right|_{(x,p)} \right),\quad
 \beta_\mu = \omega\left( \left. \frac{\partial}{\partial p^\mu} \right|_{(x,p)} \right).
\end{displaymath}

Although the adapted local coordinates $(x^\mu,p^\mu)$ are very useful, $TM$ admits many other coordinate systems whose properties may play an important role in kinetic theory. For an explicit example, we refer to section~\ref{Sec:Kerr}.

\subsection{The projection map and vertical subspace}

The map $\pi: TM\to M$ induces a projection $\pi_{*(x,p)}: T_{(x,p)}(TM)\to T_x M$ through the push-forward of $\pi$, defined as $\pi_{*(x,p)}(Z)[h] := Z[ h\circ\piÊ]$ for a tangent vector $Z$ in $T_{(x,p)}(TM)$ and a function $h: M\to \Real$ which is differentiable at $x$. It is a simple matter to verify that
\begin{equation}
\pi_{*(x,p)}\left( \left. \frac{\partial}{\partial x^\mu} \right|_{(x,p)} \right)
 = \left. \frac{\partial}{\partial x^\mu} \right|_x,\qquad
\pi_{*(x,p)}\left( \left. \frac{\partial}{\partial p^\mu} \right|_{(x,p)} \right) = 0.
\end{equation}
Therefore, the projection of an arbitrary vector field on $TM$ is given in adapted local coordinates  by
\begin{displaymath}
\pi_{*(x,p)}\left( X^\mu(x,p)\left. \frac{\partial}{\partial x^\mu} \right|_{(x,p)}
+ Y^\mu(x,p)\left. \frac{\partial}{\partial p^\mu} \right|_{(x,p)} \right) 
 = X^\mu(x,p)\left. \frac{\partial}{\partial x^\mu} \right|_x.
\end{displaymath}
At any point $(x,p)\in TM$, the \emph{vertical subspace} $V_{(x,p)}$ of $T_{(x,p)}(TM)$ is defined as
\begin{equation}
V_{(x,p)} := \ker \pi_{*(x,p)} = \{ Z\in T_{(x,p)}(TM) : \pi_{*(x,p)}(Z) = 0 \}.
\end{equation}
It is easily seen that $V_{(x,p)}$ is a $n$-dimensional subspace of $T_{(x,p)}(TM)$. In terms of adapted local coordinates $(x^\mu,p^\mu)$ it is generated by the $n$ vectors $\left. \frac{\partial}{\partial p^\mu} \right|_{(x,p)}$, $\mu = 0,1,\ldots d$.

\subsection{The connection map and horizontal subspace}

Although it was straightforward to identify at any point $(x,p)$ the vertical subspace 
$V_{(x,p)}$ this is not any longer the case for the horizontal subspace $H_{(x,p)}$. In order to provide an invariant definition of $H_{(x,p)}$ we first introduce the connection map $K$ induced by the Levi-Civita connection of the background spacetime $(M,g)$. For further properties regarding this map the reader is referred to Refs.~\cite{pD62} and~\cite{sGeK02}.

\begin{figure}[ht]
\centerline{\resizebox{9.0cm}{!}{\includegraphics{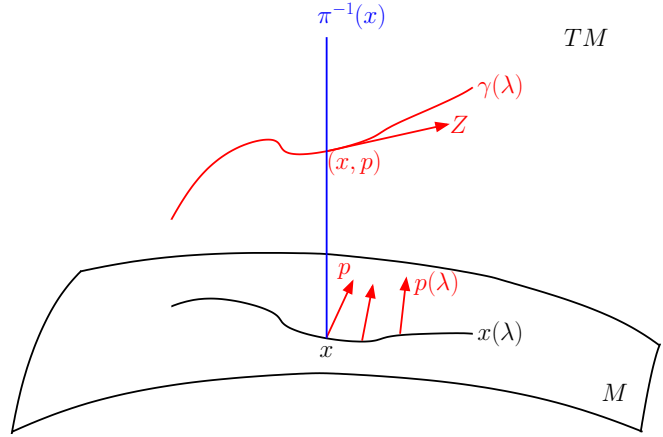}}}
\caption{The curve $\gamma(\lambda) = (x(\lambda),p(\lambda))$ on $TM$ which defines the curve $x(\lambda)$ on $M$ and the vector field $p(\lambda)$ along it.}
\label{Fig:Connection}
\end{figure}

As the push-forward of the projection map, the connection map is a linear map
\begin{equation}
K_{(x,p)}: T_{(x,p)}(TM) \to T_x M, Z\mapsto K_{(x,p)}(Z).
\end{equation}
In order to define it, let $\gamma(\lambda) = (x(\lambda),p(\lambda))$ be a smooth curve in $TM$ through $(x,p)$ with tangent vector $Z$ at $(x,p)$, that is, $\gamma(0) = (x,p)$ and $\dot{\gamma}(0) = Z$. The curve $\gamma(\lambda)$ defines the curve $x(\lambda)$ in $M$ and the vector field $p(\lambda)$ along it, see Fig.~\ref{Fig:Connection}.

By appealing to the Levi-Civita connection $\nabla$ of $(M,g)$, we define
\begin{equation}
K_{(x,p)}(Z) 
 := \left. \frac{d}{d\lambda} \tau_{0,\lambda} p(\lambda) \right|_{\lambda=0}
 = (\nabla_{\dot{x}(0)} p)_x,
\label{Eq:KDef}
\end{equation}
where $\tau_{0,\lambda}: T_{x(\lambda)} M\to T_x M$ denotes the parallel transport operator along $x(\lambda)$. Note that if $x(\lambda) = x = const.$, the parallel transport operator $\tau_{0,\lambda}$ reduces to the identity and in this case $K_{(x,p)}(Z)$ is the first variation of the family $p(\lambda)$ of tangent vectors at the point $x$. If we parametrize the curve $\gamma(\lambda)$ in terms of adapted local coordinates, $(x^\mu(\lambda),p^\mu(\lambda))$, then it follows directly from the definition that
\begin{equation}
K_{(x,p)}(Z) = \left[ \dot{p}^\mu(0) 
 + \Gamma^\mu{}_{\alpha\beta}(x)\dot{x}^\alpha(0) p^\beta \right]
 \left. \frac{\partial}{\partial x^\mu} \right|_x,
\label{Eq:KDefCoord}
\end{equation}
where here $\dot{p}^\mu(0) := \left. \frac{d}{d\lambda} p^\mu(\lambda) \right|_{\lambda=0}$, $\dot{x}^\alpha(0) := \left. \frac{d}{d\lambda} x^\alpha(\lambda) \right|_{\lambda=0}$ are the components of $Z$:
\begin{displaymath}
Z = \dot{x}^\mu(0)\left. \frac{\partial}{\partial x^\mu} \right|_{(x,p)}
 + \dot{p}^\mu(0)\left. \frac{\partial}{\partial p^\mu} \right|_{(x,p)},
\end{displaymath}
and $\Gamma^\mu{}_{\alpha\beta}$ denote the Christoffel symbols of $\nabla$. Eq.~(\ref{Eq:KDefCoord}) shows that $K_{(x,p)}$ is independent of the curve $\gamma$, it only depends on its tangent vector $Z$ at $(x,p)$. From the above definition of $K$ we can infer the following properties:

\begin{lemma}
The connection map $K_{(x,p)} : T_{(x,p)}(TM)\to T_x M$ satisfies:
\begin{enumerate}
\item[(i)] $K_{(x,p)}$ is a linear map.
\item[(ii)] $H_{(x,p)} := \ker K_{(x,p)}$ is a $n$-dimensional subspace of $T_{(x,p)}(TM)$.
\item[(iii)] $H_{(x,p)} \cap V_{(x,p)} = \{ 0 \}$.
\end{enumerate}
\end{lemma}

\proof It is sufficient to establish the lemma in adapted local coordinates $(x^\mu,p^\mu)$ in a vicinity of the point $(x,p)$. In these coordinates,
\begin{displaymath}
K_{(x,p)}(Z) = \left[ Y^\mu 
 + \Gamma^\mu{}_{\alpha\beta}(x) X^\alpha p^\beta \right]
 \left. \frac{\partial}{\partial x^\mu} \right|_x,\quad
 Z = X^\mu\left. \frac{\partial}{\partial x^\mu} \right|_{(x,p)}
 + Y^\mu\left. \frac{\partial}{\partial p^\mu} \right|_{(x,p)}.
\end{displaymath}
From this it follows directly that $K_{(x,p)}(Z_1 + \mu Z_2) = K_{(x,p)}(Z_1) + \mu K_{(x,p)}(Z_2)$ for all $\mu\in\Real$, $Z_1,Z_2\in T_{(x,p)}(TM)$. This proves (i). As for (ii), it is sufficient to notice that the $n$ tangent vectors
\begin{equation}
\left. e_\mu \right|_{(x,p)} := \left. \frac{\partial}{\partial x^\mu} \right|_{(x,p)} 
 - \Gamma^\alpha{}_{\mu\beta}(x) p^\beta
 \left. \frac{\partial}{\partial p^\alpha} \right|_{(x,p)},\qquad
\mu = 0,1,\ldots d,
\label{Eq:emu}
\end{equation}
form a basis of $\ker K_{(x,p)}$. Since the tangent vectors $\left. e_\mu \right|_{(x,p)}$ are linearly independent of the tangent vectors $\left. \frac{\partial}{\partial p^\mu} \right|_{(x,p)}$ which form a basis of $V_{(x,p)}$, (iii) follows and the lemma is proven.
\qed

\subsection{Splitting into horizontal and vertical spaces}

As a consequence of the above Lemma, it follows that at each point $(x,p)\in TM$ the tangent space $T_{(x,p)}(TM)$ of the tangent bundle at $(x,p)$ splits naturally into horizontal and vertical subspaces:
\begin{equation}
T_{(x,p)}(TM) = H_{(x,p)} \oplus V_{(x,p)}.
\label{Eq:HorVerSplit}
\end{equation}
Therefore, each tangent vector $Z\in T_{(x,p)}(TM)$ can be uniquely decomposed as
\begin{displaymath}
Z = Z^H + Z^V,\qquad Z^H\in H_{(x,p)},\quad Z^V\in V_{(x,p)}.
\end{displaymath}
In terms of the basis vectors $\{ \left. e_\mu \right|_{(x,p)} , \left. \frac{\partial}{\partial p^\mu} \right|_{(x,p)} \}$ the horizontal and vertical components can be written as
\begin{equation}
Z^H = X^\mu\left. e_\mu \right|_{(x,p)},\quad  
Z^V =  Y^\mu\left. \frac{\partial}{\partial p^\mu} \right|_{(x,p)},
\end{equation}
where $X^\mu = dx^\mu_{(x,p)}(Z)$ and $Y^\mu = \theta^\mu_{(x,p)}(Z)$ with
\begin{equation}
\theta^\mu_{(x,p)} 
 := dp^\mu_{(x,p)} + \Gamma^\mu{}_{\alpha\beta}(x) p^\beta dx^\alpha_{(x,p)},
 \qquad \mu = 0,1,\ldots d.
\label{Eq:thetamu}
\end{equation}
The covectors $\{ dx^\mu_{(x,p)}, \theta^\mu_{(x,p)} \}$ form the dual basis of $T_{(x,p)}^*(TM)$.

As a further consequence of the connection map and the push-forward of the projection map, each of the two spaces $H_{(x,p)}$ and $V_{(x,p)}$ can be identified naturally with the tangent space $T_x M$ at $x$ through the restriction of $\pi_{*(x,p)}$ to the horizontal subspace and the restriction of the connection map $K_{(x,p)}$ to the vertical subspace. Specifically, these maps are given by
\begin{eqnarray}
&& I^H_{(x,p)} := \left. \pi_{*(x,p)} \right|_{H_{(x,p)}} : H_{(x,p)} \to T_x M,
\label{Eq:IH}\\
&& I^V_{(x,p)} := \left. K_{(x,p)} \right|_{V_{(x,p)}} : V_{(x,p)} \to T_x M,
\label{Eq:IV}
\end{eqnarray}
and are linear isomorphisms. As a consequence of these maps, it is also possible to identify the horizontal and the vertical subspaces through the linear isomorphism
\begin{displaymath}
Q_{(x,p)}: H_{(x,p)} \to V_{(x,p)},
Z^H\mapsto Q(Z^H) := (I^V_{(x,p)})^{-1}\circ I^H_{(x,p)}(Z^H).
\end{displaymath}

Finally, for the development of the kinetic theory, we introduce an almost complex structure on $TM$. This is a linear map $J_{(x,p)} : T_{(x,p)}(TM) \to T_{(x,p)}(TM)$ 
which we define in terms of the map $Q_{(x,p)}$ by
\begin{equation}
J_{(x,p)}(Z) = J_{(x,p)}(Z^H + Z^V)
 := Q_{(x,p)}(Z^H) - Q_{(x,p)}^{-1}(Z^V),
\label{Eq:AlmostComplexStruc}
\end{equation}
for $Z\in T_{(x,p)}(TM)$. In a matrix notation adapted to the splitting~(\ref{Eq:HorVerSplit}) this map can also be written as
\begin{equation}
J\left( \begin{array}{c} Z^H \\ Z^V \end{array} \right)
 = \left( \begin{array}{rr} 0 & -Q^{-1} \\ Q & 0 \end{array} \right)
 \left( \begin{array}{c} Z^H \\ Z^V \end{array} \right).
\end{equation}
Therefore, it is seen that $J$ rotates the horizontal space to the vertical space and vice versa. In particular, it follows that $J^2 = -1$. In terms of the basis vectors $\{ \left. e_\mu \right|_{(x,p)} , \left. \frac{\partial}{\partial p^\mu} \right|_{(x,p)} \}$ of $T_{(x,p)}(TM)$ the maps $I^H_{(x,p)}$, $I^V_{(x,p)}$ and $J_{(x,p)}$ have the following representations:
\begin{eqnarray}
&& I^H_{(x,p)}\left( X^\mu\left. e_\mu \right|_{(x,p)} \right) = 
X^\mu\left. \frac{\partial}{\partial x^\mu} \right|_x ,
\label{Eq:IHLocCoord}\\
&& I^V_{(x,p)}\left( Y^\mu\left. \frac{\partial}{\partial p^\mu} \right|_{(x,p)} \right) = 
Y^\mu\left. \frac{\partial}{\partial x^\mu} \right|_x ,
\label{Eq:IVLocCoord}\\
&& J_{(x,p)}\left( X^\mu\left. e_\mu \right|_{(x,p)} 
 + Y^\mu\left. \frac{\partial}{\partial p^\mu} \right|_{(x,p)} \right)
 =  -Y^\mu\left. e_\mu \right|_{(x,p)} 
 + X^\mu\left. \frac{\partial}{\partial p^\mu} \right|_{(x,p)}.
\label{Eq:JLocCoord}
\end{eqnarray}
As seen from these expressions, the operators $I^H_{(x,p)}$, $I^V_{(x,p)}$, $J_{(x,p)}$ are smooth in $(x,p)$.

\section{The canonical bundle metric}
\label{Sec:SSMetric}

Based on the structures on the tangent bundle introduced in the previous section, here we define a natural metric $\hat{g}$ on $TM$. This metric, which was introduced a long time ago by Sasaki~\cite{sS58} in the context of Riemannian manifolds will be fundamental in our approach to the development of relativistic kinetic theory presented in the next section. The metric $\hat{g}$ is defined in terms of the Lorentzian metric $g$ on the base manifold and the push-forward of $\pi$ and the connection map $K$ as follows:
\begin{equation}
\hat{g}_{(x,p)}(Z,W)  := g_x( \pi_{*(x,p)}(Z), \pi_{*(x,p)}(W) ) + g_x( K_{(x,p)}(Z), K_{(x,p)}(W) ),
\label{Eq:BundleMetric}
\end{equation}
for $Z,W\in T_{(x,p)}(TM)$. The right-hand side is well-defined, bilinear and symmetric in $(Z,W)$. Furthermore, $\hat{g}_{(x,p)}$ is non-degenerated. In order to prove this, suppose $\hat{g}_{(x,p)}(Z,W) = 0$ for all $W\in T_{(x,p)}(TM)$. Choosing $W = W^H$ horizontal, it follows from Eq.~(\ref{Eq:BundleMetric}) that $g_x( \pi_{*(x,p)}(Z), \pi_{*(x,p)}(W^H) ) = 0$ for all such $W^H$ which implies that $\pi_{*(x,p)}(Z) = 0$ since $g_x$ is non-degenerated. Similarly, choosing $W = W^V$ vertical implies that $K_{(x,p)}(Z) = 0$. Consequently, $Z  = 0$ and thus the metric $\hat{g}$ is non-degenerated. For further properties of the Sasaki metric we refer the reader to~\cite{sS58,sGeK02}.

In terms of the basis vector $\{ \left. e_\mu \right|_{(x,p)} , \left. \frac{\partial}{\partial p^\mu} \right|_{(x,p)} \}$ of $T_{(x,p)}(TM)$ defined in Eq.~(\ref{Eq:emu}) the components of the metric $\hat{g}$ are:
\begin{eqnarray*}
\hat{g}_{(x,p)}\left( \left. e_\mu \right|_{(x,p)} , \left. e_\nu  \right|_{(x,p)} \right)
 &=& g_x\left( \left. \frac{\partial}{\partial x^\mu} \right|_x,
\left. \frac{\partial}{\partial x^\nu} \right|_x \right) = g_{\mu\nu}(x),
\\
\hat{g}_{(x,p)}\left( \left. \frac{\partial}{\partial p^\mu} \right|_{(x,p)},
\left. \frac{\partial}{\partial p^\nu} \right|_{(x,p)} \right) 
 &=& g_x\left( \left. \frac{\partial}{\partial x^\mu} \right|_x,
\left. \frac{\partial}{\partial x^\nu} \right|_x \right) = g_{\mu\nu}(x),\\
\hat{g}_{(x,p)}\left( \left. e_\mu \right|_{(x,p)} ,
\left. \frac{\partial}{\partial p^\nu} \right|_{(x,p)} \right) &=& 0,
\end{eqnarray*}
from which it follows that
\begin{equation}
\hat{g}_{(x,p)} = g_{\mu\nu}(x) dx^\mu_{(x,p)}\otimes dx^\nu_{(x,p)}
 + g_{\mu\nu}(x)\theta^\mu_{(x,p)}\otimes\theta^\nu_{(x,p)},
\label{Eq:BundleMetricCoord}
\end{equation}
with $\{ dx^\mu_{(x,p)}, \theta^\mu_{(x,p)} \}$ the dual basis of $T_{(x,p)}^*(TM)$, where $\theta^\mu_{(x,p)}$ is defined in Eq.~(\ref{Eq:thetamu}). As a consequence of Eq.~(\ref{Eq:BundleMetricCoord}) it is seen that the signature of $\hat{g}$ is $(2,2n-2)$. Next, we mention the following Lemma which is an immediate consequence of the definition of $\hat{g}$.

\begin{lemma}
\begin{enumerate}
\item[(i)] The splitting in Eq.~(\ref{Eq:HorVerSplit}) is orthogonal with respect to $\hat{g}$, that is, $\hat{g}_{(x,p)}(Z^H,Z^V) = 0$ for $Z^H\in H_{(x,p)}$ and $Z^V\in V_{(x,p)}$.
\item[(ii)] The metric $\hat{g}$ is invariant with respect to the almost complex structure $J$ defined in Eq.~(\ref{Eq:AlmostComplexStruc}): $\hat{g}_{(x,p)}(J(Z),J(W)) = \hat{g}_{(x,p)}(Z,W)$ for all $Z,W\in T_{(x,p)}(TM)$.
\end{enumerate}
\end{lemma}

For later use, we point out here the following property: let $\{ e_0,e_1,\ldots,e_d \}$ be a local orthonormal frame of $(M,g)$, then
\begin{equation}
\left\{ (I^H)^{-1}(e_0),\ldots, (I^H)^{-1}(e_d),(I^V)^{-1}(e_0),\ldots, (I^V)^{-1}(e_d) 
\right\}
\label{Eq:TMONB}
\end{equation}
is a local orthonormal frame of $(TM,\hat{g})$ which is adapted to the splitting~(\ref{Eq:HorVerSplit}).

Finally, we notice that the metric $\hat{g}$ in conjunction with the almost complex structure $J$ defines a natural symplectic form $\Omega_s$ by
\begin{equation}
\Omega_s(Z,W) := \hat{g}(Z, J(W))
\label{Eq:Omegas}
\end{equation}
for two vector fields $Z$ and $W$ on $TM$. It is simple to verify that $\Omega_s$ is antisymmetric and non-degenerated. In terms of the basis vector $\{ e_\mu, \frac{\partial}{\partial p^\mu} \}$ we have
\begin{eqnarray*}
\Omega_s(e_\mu,e_\nu) 
 &=& \hat{g}\left( e_\mu, \frac{\partial}{\partial p^\nu} \right) = 0,\\
\Omega_s\left(\frac{\partial}{\partial p^\mu},\frac{\partial}{\partial p^\nu} \right)
 &=& \hat{g}\left( \frac{\partial}{\partial p^\mu}, -e_\nu \right) = 0,\\
\Omega_s\left(e_\mu,\frac{\partial}{\partial p^\nu} \right) 
 &=& \hat{g}(e_\mu,-e_\nu) = -g_{\mu\nu},
\end{eqnarray*}
from which
\begin{equation}
\Omega_s = g_{\mu\nu}\theta^\mu\wedge dx^\nu 
 = g_{\mu\nu} dp^\mu\wedge dx^\nu 
 + \frac{\partial g_{\mu\nu}}{\partial x^\alpha} p^\mu dx^\alpha\wedge dx^\nu
 = d\Theta,
\label{Eq:OmegasLocCoord}
\end{equation}
where $\Theta = g_{\mu\nu} p^\mu dx^\nu$ is the Poincar\'e one-form on the tangent bundle, defined invariantly by
\begin{equation}
\Theta_{(x,p)}(X) := g_x(p,\pi_{*(x,p)}(X)),\qquad X\in T_{(x,p)}(TM).
\label{Eq:PoincareOneForm}
\end{equation}
In particular, it follows that $\Omega_s$ is closed, and thus defines a symplectic structure on $TM$. A further important property of the symplectic form that will be important in view of Liouville's theorem is the following relation with the volume form $\eta_{TM}$ induced by the bundle metric $\hat{g}$, defined as\footnote{Note that the orientation induced by this volume form differs from the one induced by the volume form $\Lambda$ defined in Ref.~\cite{oStZ13}.}
\begin{equation}
\eta_{TM} := -\det(g_{\mu\nu})
dx^0\wedge \ldots\wedge dx^d\wedge
dp^0\wedge \ldots\wedge dp^d.
\label{Eq:VolumeFormTM}
\end{equation}

\begin{lemma}
\label{Lem:VolumeTM}
Let $\eta_{TM}$ be the volume form on $TM$ induced by the bundle metric $\hat{g}$. Then,
\begin{equation}
\eta_{TM} = -\frac{(-1)^{\frac{n(n+1)}{2}}}{n!} 
\Omega_s\wedge\Omega_s\wedge\ldots\wedge\Omega_s,\qquad
\hbox{($n$-fold product)}.
\end{equation}
\end{lemma}

\proof Using the local coordinate expressions in Eqs.~(\ref{Eq:BundleMetricCoord},\ref{Eq:OmegasLocCoord}) we find
\begin{eqnarray*}
\eta_{TM} &=& -\det(g_{\mu\nu})
dx^0\wedge \ldots\wedge dx^d\wedge 
dp^0\wedge \ldots\wedge dp^d\\
 &=& -\frac{1}{n!} g_{\mu_0\nu_0}\ldots g_{\mu_d\nu_d}
\varepsilon^{\mu_0\mu_1\ldots\mu_d}\varepsilon^{\nu_0\nu_1\ldots\nu_d}
dx^0\wedge \ldots\wedge dx^d\wedge 
dp^0\wedge \ldots\wedge dp^d\\
 &=& -\frac{1}{n!} g_{\mu_0\nu_0}\ldots g_{\mu_d\nu_d}
dx^{\mu_0}\wedge dx^{\mu_1}\wedge \ldots\wedge dx^{\mu_d}\wedge
dp^{\nu_0}\wedge dp^{\nu_1}\wedge \ldots\wedge dp^{\nu_d}\\
 &=& -\frac{(-1)^k}{n!} g_{\mu_0\nu_0}\ldots g_{\mu_d\nu_d}
dp^{\mu_0}\wedge dx^{\nu_0}\wedge dp^{\mu_1}\wedge dx^{\nu_1}\wedge\ldots\wedge
dp^{\mu_d}\wedge dx^{\nu_d}\\
 &=& -\frac{(-1)^k}{n!} \Omega_s\wedge\Omega_s\wedge\ldots\wedge\Omega_s,
\end{eqnarray*}
where $k = n(n+1)/2$ is the number of transpositions used in the fourth step.
\qed

\section{The kinetic theory for a simple collisionless gas}
\label{Sec:KTheory}

After discussing the geometrical aspects of the tangent bundle, in this section we apply this framework to the description of relativistic kinetic theory for a simple gas. As we mentioned in the introduction, one possible approach to the development of the theory is based on the symplectic structure derived from the Poincar\'e one-form of $TM$ and on the introduction of a suitable Hamiltonian function, see Ref.~\cite{oStZ13}.

In the present article a complementary description is advanced which, we believe, offers new insight into the structure of relativistic kinetic theory. This new approach is based on the natural splitting of the tangent space of $TM$ into horizontal and vertical subspaces and on the associated canonical bundle metric $\hat{g}$ discussed in the previous section. In a first step, the Liouville vector field $L$ is defined as an appropriate horizontal vector field on $TM$. Next, $L$ and $\hat{g}$ define a natural Hamiltonian which, in turn, defines the mass shells $\Gamma_m$. We review some of the basic topological properties of $\Gamma_m$ that are of relevance to relativistic kinetic theory.

In contrast to the approach in~\cite{oStZ13}, now there exists a metric on each mass shell, induced by $\hat{g}$. As we show, this turns the mass shells into Lorentzian manifolds on which a natural integration theory can be developed. The integrals solely involve the metric structure instead of the volume form induced by the symplectic form. We show  that the Liouville vector field $L$ is divergence-free and thus it generates an incompressible flow on $\Gamma_m$. Next, we introduce the distribution function $f$ on the mass shell. The metric $\hat{g}$ in combination with the Liouville vector field $L$ allows us to advance a new interpretation of the distribution function $f$. For this, $f$ together with $L$ are thought of as defining a fictitious incompressible fluid on $\Gamma_m$ with current density ${\cal J} = f L/m$, where $f$ plays the analogue of the particle density. Accordingly, flux integrals of ${\cal J}$ over suitable spacelike hypersurfaces $\Sigma$ in $\Gamma_m$ are interpreted as ensemble averages of the number of occupied trajectories intersecting $\Sigma$.

The present definition of the distribution function, although different from the one used in~\cite{oStZ13}, leads to equivalent observables.

\subsection{The Liouville vector field}

We start with the definition of the Liouville vector field $L\in {\cal X}(TM)$. For this, we recall the linear, invertible map $I^H_{(x,p)}: H_{(x,p)}\to T_x M$ in Eq.~(\ref{Eq:IH}) and define
\begin{equation}
L_{(x,p)} := (I^H_{(x,p)})^{-1}(p), \qquad (x,p)\in TM.
\label{Eq:LDef}
\end{equation}
By definition, $L_{(x,p)}$ is the unique tangent vector of $TM$ at $(x,p)$ which is horizontal, that is, $K_{(x,p)}(L_{(x,p)}) = 0$, and which satisfies $\pi_{*(x,p)}(L_{(x,p)}) = p$. As a consequence, any integral curve $(x(\lambda),p(\lambda))$ of $L$ through the point $(x,p)$ obeys 
\begin{displaymath}
\dot{x}(0) = p,\qquad  \nabla_{\dot{x}(0)} p = 0.
\end{displaymath}
Therefore, the projected curve $x(\lambda)$ on $M$ defines a geodesic of the base manifold $(M,g)$. It is natural to ask whether the Liouville vector field $L$ also generates geodesics on the tangent bundle $(TM,\hat{g})$. This turns out to be the case as formulated in the following Proposition.

\begin{proposition}[cf.~\cite{sS58}]
\label{Prop:LGeodesic}
The Liouville vector field $L$ satisfies the geodesic equation
\begin{equation}
\hat{\nabla}_L L = 0
\end{equation}
on $(TM,\hat{g})$, where $\hat{\nabla}$ denotes the Levi-Civita connection with respect to $\hat{g}$.
\end{proposition}

\proof See Ref.~\cite{sS58} or the appendix for a proof which avoids computing explicitly the connection $\hat{\nabla}$.
\qed

In terms of adapted local coordinates the Liouville vector field reads
\begin{equation}
L_{(x,p)} = p^\mu \left. e_\mu \right|_{(x,p)} 
 = p^\mu\left. \frac{\partial}{\partial x^\mu} \right|_{(x,p)} 
 - \Gamma^\mu{}_{\alpha\beta}(x) p^\alpha p^\beta
 \left. \frac{\partial}{\partial p^\mu} \right|_{(x,p)},
\label{Eq:LLocCoord}
\end{equation}
where we have used Eqs.~(\ref{Eq:IHLocCoord}) and (\ref{Eq:emu}). 

In the context of relativistic kinetic theory the integral curves of $L$ corresponding to timelike geodesics describe the possible trajectories of gas particles.

\subsection{Hamiltonian and mass shell}

Next, we introduce the Hamiltonian function $H: TM\to \Real$, defined by
\begin{equation}
H := \frac{1}{2}\hat{g}(L,L).
\label{Eq:HDef}
\end{equation}
Using the definitions of $\hat{g}$ and the fact that $L$ is horizontal, we obtain
\begin{equation}
H(x,p) = \frac{1}{2} g_x\left( \pi_{*(x,p)}(L_{(x,p)}), \pi_{*(x,p)}(L_{(x,p)}) \right)
 = \frac{1}{2} g_x(p,p).
\label{Eq:HDefBis}
\end{equation}

Once a Hamiltonian function has been defined, then for any $m$ we define
the mass shell via 
\begin{equation}
\Gamma_m := H^{-1}\left( -\frac{1}{2}m^{2} \right)
 = \{ (x,p)\in TM : g_x(p,p) = -m^2 \}.
\end{equation} 

\begin{lemma}
\label{Lem:MassShell}
For any $m\neq 0$ the set $\Gamma_m$ defines a $(2n-1)$-dimensional $C^\infty$-differentiable manifold.
\end{lemma}

\proof Let us compute the differential of $H$ in adapted local coordinates $(x^\mu,p^\mu)$, and show that it is different from zero at points $(x,p)\in \Gamma_m$ with $m\neq 0$. Using Eq.~(\ref{Eq:HDefBis}) we obtain
\begin{equation}
dH = \frac{1}{2} d\left( g_{\mu\nu} p^\mu p^\nu \right)
 = g_{\mu\nu} p^\mu dp^\nu
 + \frac{1}{2}\frac{\partial g_{\mu\nu}}{\partial x^\alpha} p^\mu p^\nu dx^\alpha
 = g_{\mu\nu} p^\mu \theta^\nu,
\label{Eq:dH}
\end{equation}
where $\theta^\nu$ is defined in Eq.~(\ref{Eq:thetamu}). Since $g_x(p,p) = -m^2 < 0$ it follows that $p\neq 0$ and thus $dH_{(x,p)} \neq 0$. Therefore, $\Gamma_m$ is a submanifold of $TM$ of co-dimension one, and the lemma follows. 
\qed

{\bf Remark}: For the massless case $m=0$ the set $\Gamma_m$ is smooth except at the vertex points $p=0$.\\

For further insights regarding the structure of $\Gamma_m$ we introduce the following subset of the tangent space $T_x M$:
\begin{equation}
P_x := \{ p\in T_x M : g_x(p,p) = -m^2 \},
\label{Eq:DefPx}
\end{equation}
in terms of which $\Gamma_m$ can also be written as $\Gamma_m = \{ (x,p) : x\in M, p\in P_x \}$. For an arbitrary spacetime $(M,g)$, $P_x$ is the union of two disjoint sets $P_x^+$ and $P_x^-$, which may be called the ``future'' and the ``past'' mass hyperboloid at $x$, respectively. Because the spacetime metric is smooth, this distinction can be extended unambiguously to a small neighborhood of $x$. However, it can be extended unambiguously to the whole spacetime only if $(M,g)$ is time-orientable. The following result connects the time-orientability of the base manifold $(M,g)$ to the topological properties of $\Gamma_m$:

\begin{proposition}
\label{Prop:TimeOrientability}
Suppose $M$ is connected and $m > 0$. Then, $(M,g)$ is time-orientable if and only if $\Gamma_m$ is disconnected, in which case it is the disjoint union of two connected components $\Gamma_m^+$ and $\Gamma_m^-$.
\end{proposition}

\proof See Appendix A in Ref.~\cite{oStZ13}.
\qed

In the following, we assume that $(M,g)$ is time-oriented and work on the ``future'' mass shell $\Gamma_m^+$. Physically, this incorporates the idea that the gas particles move on future directed timelike geodesics.

We finish this subsection by mentioning an alternative definition for the Liouville vector field $L$, employed in~\cite{oStZ13}, in which $L$ is defined as the Hamiltonian vector field belonging to the Hamiltonian $H$  on the symplectic manifold $(TM,\Omega_s)$. According to this definition,
\begin{equation}
dH = -i_L\Omega_s = \Omega_s(\cdot,L).
\label{Eq:LHam}
\end{equation}
The fact that the two definitions agree with each other follows from the following Lemma.

\begin{lemma}
Let $L$ be the Liouville vector field defined by Eq.~(\ref{Eq:LDef}). Then $L$ satisfies Eq.~(\ref{Eq:LHam}), with $\Omega_s$ the symplectic form defined in Eq.~(\ref{Eq:Omegas}).
\end{lemma}

\proof Eq.~(\ref{Eq:LHam}) can be verified either using the expressions (\ref{Eq:OmegasLocCoord},\ref{Eq:LLocCoord},\ref{Eq:dH}) in adapted local coordinates, or by the following, coordinate-independent argument: let $(x,p)\in TM$ and $Z\in T_{(x,p)}(TM)$, and let $\gamma(\lambda) = (x(\lambda),p(\lambda))$ be a smooth curve through $(x,p)$ with tangent vector $Z$ at $\lambda=0$. Then, using the definition of $\Omega_s$ in Eq.~(\ref{Eq:Omegas}) and the fact that $J(L) = (I^V)^{-1}(p)$ is vertical, we find
\begin{eqnarray*}
\Omega_{s(x,p)}(Z,L) &=& g_x( K_{(x,p)}(Z), K_{(x,p)}\circ (I^V_{(x,p)})^{-1}(p) )\\
 &=& g_x( K_{(x,p)}(Z) , p) \\
 &=& \left. \frac{d}{d\lambda} \right|_{\lambda=0}
  \frac{1}{2} g_{x(\lambda)}(p(\lambda),p(\lambda))\\
 &=& \left. \frac{d}{d\lambda} \right|_{\lambda=0} H(x(\lambda),p(\lambda))\\
 &=& dH_{(x,p)}(Z),
\end{eqnarray*}
where in the third step we have used the definition~(\ref{Eq:KDef}) of $K_{(x,p)}$ and the fact that the connection is metric compatible.
\qed

\subsection{The mass shell as a Lorentz manifold}

Since $\Gamma_m$ is a submanifold of $TM$, the canonical bundle metric $\hat{g}$ induces naturally a metric $\hat{h}$ on $\Gamma_m$. It is defined as the pull-back of $\hat{g}$ with respect to the inclusion map $\iota: \Gamma_m\to TM$, that is
\begin{equation}
\hat{h} := \iota^*\hat{g}.
\end{equation}

\begin{lemma}
\label{Lem:MassShellLM}
$(\Gamma_m,\hat{h})$ is a $(2n-1)$-dimensional Lorentz manifold with unit normal vector field $N = (I^V)^{-1}(p/m)$.
\end{lemma}

\proof Let $(x,p)\in \Gamma_m$ and consider an orthonormal basis of $T_x M$ of the form $\{ p/m, e_1,e_2,\ldots,e_d \}$. It induces on $T_{(x,p)}(TM)$ the orthonormal basis
\begin{eqnarray*}
&&
\left\{ (I^H_{(x,p)})^{-1}(p/m),(I^H_{(x,p)})^{-1}(e_1),\ldots, (I^H_{(x,p)})^{-1}(e_d),
\right.\\
&& \left. \quad 
(I^V_{(x,p)})^{-1}(p/m),(I^V_{(x,p)})^{-1}(e_1),\ldots, (I^V_{(x,p)})^{-1}(e_d) 
\right\}.
\end{eqnarray*}
By definition, the first vector is equal to $L_{(x,p)}/m$. Next, notice that $N_{(x,p)} := (I^V_{(x,p)})^{-1}(p/m) = J_{(x,p)}(L_{(x,p)})/m$. We claim that $N$ is a unit normal vector field to $\Gamma_m$. For this, let $Z$ be a vector field on $TM$. Then, on $\Gamma_m$, it follows that
\begin{equation}
m\hat{g}(N,Z) = \hat{g}( J(L), Z) = -\hat{g}( L, J(Z) ) = -\Omega_s(L,Z) = dH(Z),
\label{Eq:NdH}
\end{equation}
where we have used Eq.~(\ref{Eq:LHam}) in the last step. In particular, the right-hand side is zero if $Z$ is tangent to $\Gamma_m$, which proves that $N$ is normal to $\Gamma_m$.

It follows from Lemma~\ref{Lem:MassShell} that the vectors
\begin{displaymath}
\left\{ L_{(x,p)}/m,(I^H_{(x,p)})^{-1}(e_1),\ldots, (I^H_{(x,p)})^{-1}(e_d),(I^V_{(x,p)})^{-1}(e_1),\ldots, (I^V_{(x,p)})^{-1}(e_d) 
\right\}
\end{displaymath}
form an orthonormal basis of $T_{(x,p)}(\Gamma_m)$. The first vector is timelike with respect to $\hat{h}$; all other vectors are spacelike. Therefore, the signature of $\hat{h}$ is $(1,2n-2)$ and the lemma is shown.
\qed

{\bf Remark}: In particular, it follows from Lemma~\ref{Lem:MassShellLM} that any horizontal vector field is tangent to $\Gamma_m$.\\

For later purposes, it is useful to have a local representation of the metric $\hat{h}$. To this end, we first construct a local coordinate chart on the mass shell $\Gamma_m$. For that, let $(U,\phi)$ be a local chart of $(M,g)$ with corresponding local coordinates $(x^0,x^1,\ldots,x^d)$, such that
\begin{displaymath}
\left\{ \left. \frac{\partial}{\partial x^0} \right|_x,
\left. \frac{\partial}{\partial x^1} \right|_x,\ldots,
\left. \frac{\partial}{\partial x^d} \right|_x \right\},\qquad
x\in U,
\end{displaymath}
is a basis of $T_x M$, with the property that for each $x\in U$, $\left. \frac{\partial}{\partial x^0} \right|_x$ is timelike and all the vectors of the form $\left. p^i\frac{\partial}{\partial x^i} \right|_x$ are spacelike. Let $(V,\psi)$ denote the local chart of $TM$ with the corresponding adapted local coordinates $(x^\mu,p^\mu)$ constructed in the proof of Lemma~\ref{Lem:TM}. Relative to these local coordinates, the mass shell is determined by
\begin{equation}
-m^2 = g_{\mu\nu}(x) p^\mu p^\nu 
 = g_{00}(x)(p^0)^2 + 2 g_{0j}(x) p^0 p^j + g_{ij}(x) p^i p^j.
\label{Eq:MassShellCond}
\end{equation}
Therefore, the mass shell $\Gamma_m$ can be locally represented as those $(x^\mu,p^0,p^i)\in \psi(V)\subset \Real^{2n}$ for which $p^0 = p_\pm^0(x^\mu,p^i)$ with
\begin{equation}
p_\pm^0(x^\mu,p^i) := \frac{g_{0j}(x) p^j \pm
 \sqrt{[g_{0j}(x) p^j]^2 + [-g_{00}(x)]\left[ m^2 + g_{ij}(x) p^i p^j \right] }}
 {-g_{00}(x)}.
\label{Eq:p0up}
\end{equation}
Since $g_{00}(x) < 0$ and $g_{ij}(x) p^i p^j\geq 0$ for all $x\in U$, it follows that 
$p_+^0(x^\mu,p^i)$ is positive and $p_-^0(x^\mu,p^i)$ is negative. This liberty in the choice of $p^0$ expresses the fact that locally, $\Gamma_m$ has two disconnected components, representing ``future'' and ``past''. Since we assume $(M,g)$ to be time-oriented, this distinction can be made globally and $p_\pm^0(x^\mu,p^i)$ parametrize locally the two disconnected components $\Gamma_m^\pm$ of the mass shell, see Proposition~\ref{Prop:TimeOrientability}.

By differentiating both sides of Eq.~(\ref{Eq:MassShellCond}) one finds
\begin{displaymath}
\frac{\partial p_\pm^0(x^\mu,p^j)}{\partial x^\mu} 
 = -\frac{1}{2p_{\pm 0}(x^\mu,p^j)} \hat{p}^\alpha \hat{p}^\beta
 \frac{\partial g_{\alpha\beta}}{\partial x^\mu},\qquad
\frac{\partial p_\pm^0(x^\mu,p^j)}{\partial p^i} 
 = -\frac{1}{p_{\pm 0}(x^\mu,p^j)}\hat{p}_i,
\end{displaymath}
where here we have defined $\hat{p}^0 := p_\pm^0(x^\mu,p^j)$, $\hat{p}^j := p^j$, $\hat{p}_i := g_{i0}(x)\hat{p}^0 + g_{ij}(x)\hat{p}^j$, and 
\begin{eqnarray}
\hat{p}_0 := p_{\pm 0}(x^\mu,p^i) &:=& g_{00}(x) p_\pm^0(x^\mu,p^i) + g_{0j}(x) p^j
\nonumber\\
 &=& \mp \sqrt{ [g_{0j}(x) p^j]^2 + [-g_{00}(x)]\left[ m^2 + g_{ij}(x) p^i p^j \right]}.
\label{Eq:p0down}
\end{eqnarray}
It follows from this and Eq.~(\ref{Eq:thetamu}) that
\begin{displaymath}
\iota^* \theta^0 = -\frac{\hat{p}_k}{\hat{p}_0} \hat{\theta}^k,\qquad
\hat{\theta}^k := \iota^*\theta^k 
 = dp^k + \Gamma^k{}_{\alpha\beta}\hat{p}^\beta dx^\alpha.
\end{displaymath}
Therefore, the induced metric $\hat{h}$ in the local coordinates $(x^\mu, p^i)$ has the form
\begin{equation}
\hat{h} = g_{\mu\nu} dx^\mu\otimes dx^\nu 
 + \left( g_{ij} - \frac{2}{\hat{p}_0} g_{0(i}\hat{p}_{j)} 
  + \frac{1}{\hat{p}_0^2} g_{00}\hat{p}_i\hat{p}_j\right)
\hat{\theta}^i\otimes\hat{\theta}^j.
\end{equation}

The volume form $\eta_{\Gamma_m}$ on $\Gamma_m$ induced by the metric $\hat{h}$ is conveniently computed from
\begin{equation}
\eta_{\Gamma_m} = (-1)^n\iota^*(i_N\eta_{TM}),
\label{Eq:VolumeFormMassShell}
\end{equation}
with $\eta_{TM}$ the volume form of $(TM,\hat{g})$, $N$ the unit normal vector field to $\Gamma_m$ and $\iota: \Gamma_m\to TM$ the inclusion map. In adapted local coordinates we have
\begin{displaymath}
N = (I^V)^{-1}\left( \frac{p}{m} \right)
   = \frac{1}{m} p^\mu\frac{\partial}{\partial p^\mu}
\end{displaymath}
(see Lemma~\ref{Lem:MassShellLM}) and
\begin{displaymath}
\eta_{TM} = -\det(g_{\mu\nu})
dx^0\wedge dx^1\wedge \ldots\wedge dx^d\wedge 
dp^0\wedge dp^1\wedge \ldots\wedge dp^d.
\end{displaymath}
Using these expressions a straightforward calculation reveals that
\begin{equation}
\eta_{\Gamma_m} = \frac{m}{\hat{p}_0}\det(g_{\mu\nu})
dx^0\wedge dx^1\wedge\ldots\wedge dx^d\wedge
dp^1\wedge \ldots\wedge dp^d.
\label{Eq:VolumeFormMassShellLocalCoord}
\end{equation}
Here, we have included the factor $(-1)^n$ in the definition of $\eta_{\Gamma_m}$ so that the local coordinates $(x^\mu,p^i)$ form a positive system with respect to the orientation of $(\Gamma_m^+,N)$ induced by the natural orientation of $TM$. This volume form, when restricted to the future mass shell, agrees (up to a factor $m$) with the volume form $\Omega$ constructed in Ref.~\cite{oStZ13}.

For the formulation of the next result, it is important to note that the Liouville vector field $L\in {\cal X}(TM)$, being horizontal, is tangent to $\Gamma_m$. Therefore, we may also regard $L$ as a vector field on $\Gamma_m$.

\begin{theorem}[Liouville theorem, cf. \cite{sS62},\cite{jE73},\cite{mT85}]
\label{Thm:Liouville}
The Liouville vector field $L$, when restricted to $\Gamma_m$, satisfies
\begin{equation}
\divrg L = 0,
\end{equation}
where the divergence operator refers to the Lorentz manifold $(\Gamma_m,\hat{h})$.
\end{theorem}

\proof We first show that $L$, as a vector field on the full tangent bundle $TM$, is divergence-free with respect to $(TM,\hat{g})$. For this, we note that this is equivalent to showing
\begin{equation}
\pounds_L\eta_{TM} = 0,
\label{Eq:LiouvilleTM}
\end{equation}
with $\eta_{TM}$ the volume element on $TM$ induced by $\hat{g}$, see Eq.~(\ref{Eq:VolumeFormTM}). This in turn follows immediately from Lemma~\ref{Lem:VolumeTM} and the property that $L$ is a Hamiltonian vector field, Eq.~(\ref{Eq:LHam}), which implies $\pounds_L\Omega_s = 0$.

Next, we consider the restriction of $L$ onto the mass shell $\Gamma_m$ and show that $\pounds_L\eta_{\Gamma_m} = 0$, with $\eta_{\Gamma_m}$ the induced volume element defined in Eq.~(\ref{Eq:VolumeFormMassShell}). For this, let $X_2,\ldots X_{2n}$ be $2n-1$ vector fields on $TM$ which are tangent to $\Gamma_m$. Then, according to Eq.~(\ref{Eq:LiouvilleTM}) we have
\begin{eqnarray*}
0 &=& (\pounds_L\eta_{TM})(N,X_2\ldots,X_{2n})\\
 &=& L[ \eta_{TM}(N,X_2,\ldots,X_{2n}) ] - \eta_{TM}(\pounds_L N, X_2,\ldots,X_{2n})
 \\
 &-& \eta_{TM}(N,\pounds_L X_2,X_3,\ldots,X_{2n}) - \ldots 
 - \eta_{TM}(N,X_2,\ldots,X_{2n-1},\pounds_L X_{2n})\\
 &=& L[ \eta_{\Gamma_m}(X_2,\ldots,X_{2n}) ] 
 - \eta_{TM}(\pounds_L N, X_2,\ldots,X_{2n})\\
 &-& \eta_{\Gamma_m}(\pounds_L X_2,X_3,\ldots,X_{2n}) - \ldots 
 - \eta_{\Gamma_m}(X_2,\ldots,X_{2n-1},\pounds_L X_{2n})\\
 &=& (\pounds_L\eta_{\Gamma_m})(X_2,\ldots,X_{2n})
 - \eta_{TM}(\pounds_L N, X_2,\ldots,X_{2n}).
\end{eqnarray*}
In order to conclude the proof it is sufficient to show that $\pounds_L N$ is tangent to $\Gamma_m$, that is $\hat{g}(\pounds_L N,N) = 0$. Using Eq.~(\ref{Eq:NdH}) we find
\begin{displaymath}
m\hat{g}(\pounds_L N,N) = dH(\pounds_L N) = L[dH(N)] - (\pounds_L dH)(N).
\end{displaymath}
The first term on the right-hand side vanishes since $dH(N) = m\hat{g}(N,N) = -m$, which is constant along $L$. The second term on the right-hand side also vanishes by virtue of Eq.~(\ref{Eq:LHam}).
\qed

\subsection{Distribution function and associated current density}

As we have mentioned earlier on, for a simple collisionless gas, the world lines of the gas particles are described by integral curves of the Liouville vector field $L$ on the future mass shell $\Gamma_m^+$. For the following, we assume spacetime $(M,g)$ to be oriented and time-oriented. Following Ehlers~\cite{jE71}, we consider a Gibbs ensemble of the gas on a fixed spacetime $(M,g)$. The central assumption of relativistic kinetic theory is that the averaged properties of the gas are described by a one-particle distribution function.

\begin{figure}[ht]
\centerline{\resizebox{4.5cm}{!}{\includegraphics{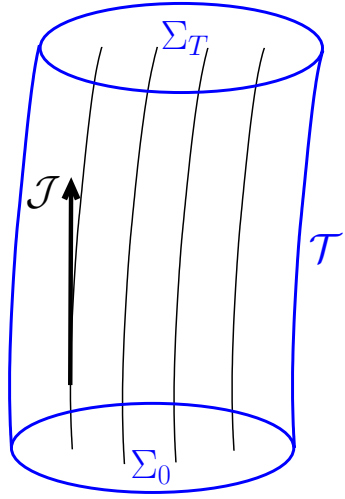}}}
\caption{The tubular region $V$ obtained by the flow of the spacelike compact hypersurface $\Sigma_0$ along ${\cal J}$. By construction ${\cal J}$ is tangent to ${\cal T}$.}
\label{Fig:Tube}
\end{figure}

Based on the properties of the mass shell discussed so far, we are now ready to explain our new interpretation of the distribution function $f$. For this, let $f\in {\cal F}(\Gamma_m^+)$ and define the associated current density
\begin{equation}
{\cal J} := f \frac{L}{m}\in {\cal X}(\Gamma_m^+).
\label{Eq:JCurrent}
\end{equation}
Notice that $\hat{h}(L/m,L/m) = -1$, so that $L/m$ plays the role of a velocity vector field on the Lorentzian manifold $(\Gamma_m,\hat{h})$. In close analogy with the fluid case, we advance the following physical interpretation for the current density ${\cal J}$ and the distribution function $f$: Given a smooth, $2d$-dimensional spacelike hypersurface $\Sigma$ in $\Gamma_m^+$ with unit normal vector field $\nu$, the flux integral
\begin{equation}
N[\Sigma] := -\int\limits_\Sigma \hat{h}({\cal J},\nu) \eta_\Sigma
\label{Eq:NSigma}
\end{equation}
gives the ensemble average of occupied trajectories that intersect $\Sigma$, where here $\eta_\Sigma$ is the volume element on $\Sigma$ induced by the metric $\hat{h}$.\footnote{The minus sign comes from the timelike character of ${\cal J}$ and $\nu$, implying that $\hat{h}({\cal J},\nu)$ is negative if they point in the same component of the light cone. Since $\eta_\Sigma = \iota^*(i_\nu\eta_{\Gamma_m^+})$ with $\iota: \Sigma\to \Gamma_m^+$ the inclusion map, by taking into account that $-\hat{h}({\cal J},\nu)\nu$ is the component of ${\cal J}$ normal to $\nu$, we can rewrite $-\hat{h}({\cal J},\nu)\eta_{\Sigma} = \iota^*(i_{\cal J}\eta_{\Gamma_m^+}) = f \iota^*(i_L\eta_{\Gamma_m^+})/m$, showing that our definition of the distribution function in Eqs.~(\ref{Eq:JCurrent},\ref{Eq:NSigma}) is equivalent to previous definitions, see for example~\cite{jE71,oStZ13}.} To understand the consequences of Eq.~(\ref{Eq:NSigma}), we consider a tubular region $V = \bigcup_{0\leq t \leq T}\Sigma_t$ which is obtained by letting flow along the integral curves of ${\cal J}$ a $2d$-dimensional, spacelike hypersurface $\Sigma_0$ in $\Gamma_m^+$. The boundary of $V$ consists of the initial and final hypersurfaces $\Sigma_0$ and $\Sigma_T$ and the cylindrical piece, ${\cal T} := \bigcup_{0\leq t\leq T}\partial\Sigma_t$, see Fig.~\ref{Fig:Tube}. By appealing to Gauss' theorem we obtain
\begin{equation}
N[\Sigma_T] - N[\Sigma_0] = \int\limits_V (\divrg{\cal J}) \eta_{\Gamma_m},
\label{Eq:BalanceLaw}
\end{equation}
where here $\eta_{\Gamma_m}$ is the volume element on $\Gamma_m^+$ induced by $\hat{h}$ and where we have used the fact that ${\cal J}$ is tangent to ${\cal T}$. The expression on the left-hand side of this equation is equal to the ensemble average of the net change in number of occupied trajectories between $\Sigma_0$ and $\Sigma_T$ due to collisions.

For the particular case of a collisionless gas it follows from Eq.~(\ref{Eq:BalanceLaw}) that $\divrg{\cal J} = 0$. Taking into account the definition of ${\cal J}$ and Liouville's theorem, $\divrg L = 0$ (see Theorem~\ref{Thm:Liouville}), this implies that the distribution function $f$ must satisfy the Liouville equation
\begin{equation}
\pounds_L f = 0.
\label{Eq:LiouvilleEq}
\end{equation}

Therefore, the distribution function $f$ and the associated current density ${\cal J} = f L/m$ describe a fictitious incompressible fluid on the future mass shell $(\Gamma_m^+,\hat{h})$, and they play the analogous role as the particle density $n$ and the particle current density associated to an incompressible fluid on $(M,g)$, for which the continuity equation implies that $n$ is constant along the flow lines.

As we show now, the current density ${\cal J}$ on $TM$ gives rise to a natural current $J$ on the spacetime which is divergence-free. In order to explain this property we consider a $d$-dimensional spacelike hypersurface  $S$ in $(M,g)$ with unit normal vector field $s$. This surface can be lifted to a $2d$-dimensional hypersurface $\Sigma$ of $\Gamma_m^+$, defined as
\begin{equation}
\Sigma := \{ (x,p) : x\in S, p\in P_x^+ \} \subset \Gamma_m^+
\label{Eq:Sigma}
\end{equation}
with associated unit normal vector field $\nu := (I^H)^{-1}(s)$. In order to verify that $\nu$ is a unit normal vector field to $\Sigma$, we first note that $\nu$, like any horizontal vector field, is tangent to $\Gamma_m^+$. Next, for each $Z\in {\cal X}(TM)$ we have
\begin{displaymath}
\hat{g}(\nu,Z) = g(s,\pi_*(Z))
\end{displaymath}
since $\nu$ is horizontal. In particular, for $Z=\nu$ it follows that $\hat{g}(\nu,\nu) = g(s,s) = -1$ which shows that $\nu$ is normalized. Finally, let $Z$ be the tangent vector to any smooth curve $\gamma(\lambda) = (x(\lambda),p(\lambda))$ in $\Sigma$. Then,
\begin{displaymath}
\pi_*(Z) = \frac{d}{d\lambda} \pi(\gamma(\lambda)) = \frac{d}{d\lambda} x(\lambda),
\end{displaymath}
which is orthogonal to $s$ since $x(\lambda)\in S$. Therefore, $\hat{g}(\nu,Z) = 0$ and thus $\nu$ is normal to $\Sigma$.

Now let us compute the ensemble average of occupied trajectories that intersect a hypersurface $\Sigma$ of the form~(\ref{Eq:Sigma}). Assuming for simplicity that $S$ is entirely contained inside a coordinate neighborhood of $M$ we have, in adapted local coordinates,
\begin{displaymath}
\nu = (I^H)^{-1}(s) = s^\mu e_\mu,
\end{displaymath}
and together with the expression~(\ref{Eq:VolumeFormMassShellLocalCoord}) this yields
\begin{eqnarray}
i_\nu\eta_{\Gamma_m} &=& \frac{m}{\hat{p}_0}\det(g_{\mu\nu})
\frac{1}{d!} s^{\mu_0}\varepsilon_{\mu_0\mu_1\ldots\mu_d} 
dx^{\mu_1}\wedge\ldots\wedge dx^{\mu_d}\wedge dp^1\wedge\ldots\wedge dp^d
\nonumber\\
 &=& m(i_s\eta)\wedge \pi_x,
\label{Eq:LocalSplitting}
\end{eqnarray}
where we have introduced the natural volume element $\eta$ on $(M,g)$, defined by
\begin{displaymath}
\eta := \sqrt{-\det(g_{\mu\nu})} dx^0\wedge dx^1\wedge\ldots\wedge dx^d,
\end{displaymath}
and the natural volume element on $P_x^+$, defined by
\begin{displaymath}
\pi_x := \frac{\sqrt{-\det(g_{\mu\nu})}}{-\hat{p}_0} dp^1\wedge\ldots\wedge dp^d.
\end{displaymath}
This volume element arises naturally by considering the induced metric on the future mass hyperboloid $P_x^+$, defining $\pi_x$ as the associated volume form. Using the local splitting (\ref{Eq:LocalSplitting}), it follows that the integral of any sufficiently smooth and fast decaying function $F$ over a hypersurface $\Sigma$ of the form~(\ref{Eq:Sigma}) can be computed as
\begin{equation}
\int\limits_\Sigma F \eta_\Sigma 
 = m\int\limits_S \left( \int\limits_{P_x^+} F \pi_x \right) \eta_S,
\end{equation}
with $\eta_S := i_s\eta$ the induced volume element on $S$. This Fubini-type formula still holds true if the surface $S$ is not included in a coordinate chart, as can be shown by using a partition of unity. It also holds for timelike hypersurfaces, where $s$ is spacelike. In particular for $F(x,p) = -\hat{h}_{(x,p)}({\cal J},\nu) = -f(x,p) g_x(p,s)/m$ it follows that
\begin{equation}
N[\Sigma] = -\int\limits_\Sigma \hat{h}({\cal J},\nu) \eta_\Sigma
 = -\int\limits_S g(J,s) \eta_S,
\label{Eq:JJ}
\end{equation}
where the current density $J\in {\cal X}(M)$ is defined as
\begin{equation}
J_x(\omega) := \int\limits_{P_x^+} f(x,p) \omega(p) \pi_x,\qquad
\omega\in T_x^*M,
\end{equation}
or in adapted local coordinates,
\begin{equation}
J_x^\mu = \int\limits_{P_x^+} f(x,p) p^\mu \pi_x.
\end{equation}
This vector field $J$ can be interpreted as the first moment of the distribution function in momentum space. Its flux integral through the hypersurface $S$ is equal to the flux integral of ${\cal J}$ through $\Sigma$ which is equal to $N[\Sigma]$. Obviously however, the vector field ${\cal J}$ on $\Gamma_m^+$ contains much more information than $J$ since it allows one to compute $N[\Sigma]$ for arbitrary $2d$-dimensional spatial hypersurfaces of $\Gamma_m^+$ and not just those which are of the form~(\ref{Eq:Sigma}).

The next result shows that the divergence of $J$ is also related to the divergence of ${\cal J}$ through a fibre integral, and implies that $J$ is conserved if $\divrg {\cal J} = 0$.

\begin{proposition}
\label{Prop:ConservationLawJ}
Let $f: \Gamma_m^+\to\Real$ be a $C^\infty$-function of compact support on the future mass shell. Then, the following identity holds for all $x\in M$:
\begin{equation}
\divrg J_x = m\int\limits_{P_x^+} \divrg{\cal J}(x,p)\pi_x.
\label{Eq:DivJ}
\end{equation}
\end{proposition}

\proof Consider a compact subset $K\subset M$ of $M$ with $C^\infty$-boundary $S := \partial K$ and the corresponding subset
\begin{displaymath}
V := \{ (x,p) : x\in K, p\in P_x^+ \}
\end{displaymath}
of $\Gamma_m^+$ with boundary
\begin{displaymath}
\Sigma := \partial V = \{ (x,p) : x\in S, p\in P_x^+ \}.
\end{displaymath}
Using Eq.~(\ref{Eq:JJ}) and Gauss' theorem we obtain on one hand
\begin{displaymath}
N[\Sigma] = \int\limits_K (\divrg J) \eta,
\end{displaymath}
and on the other hand using Eq.~(\ref{Eq:BalanceLaw}),
\begin{displaymath}
N[\Sigma] = \int\limits_V (\divrg{\cal J}) \eta_{\Gamma_m}
 = m\int\limits_K \left( \int\limits_{P_x^+} (\divrg{\cal J})\pi_x \right) \eta,
\end{displaymath}
where we have used the fact that in adapted local coordinates $\eta_{\Gamma_m} = m\eta\wedge \pi_x$. The statement of the proposition follows by comparing the two expressions for $N[\Sigma]$ with each other.
\qed

Since $m\divrg({\cal J}) = \divrg( f L) = \pounds_L f$, we can also rewrite Eq.~(\ref{Eq:DivJ}) as
\begin{equation}
\divrg J_x = \int\limits_{P_x^+} (\pounds_L f)(x,p)\pi_x,
\label{Eq:DivJBis}
\end{equation}
implying that $J$ is conserved if the distribution function $f$ satisfies the Liouville equation $\pounds_L f = 0$. This formula can also be generalized to higher moments of the distribution function: let $T$ be the $s$-rank contravariant symmetric tensor field on $M$ defined by
\begin{equation}
T_x(\omega^1,\ldots,\omega^s)
 := \int\limits_{P_x^+} f(x,p) \omega^1(p)\cdot\cdot\cdot\omega^s(p) \pi_x,
\qquad \omega^1,\ldots,\omega^s\in T_x^*M,
\end{equation}
or in adapted local coordinates:
\begin{equation}
T_x^{\mu_1\mu_2,\ldots,\mu_s}
 = \int\limits_{P_x^+} f(x,p) p^{\mu_1} p^{\mu_2}\cdot\cdot\cdot p^{\mu_s} \pi_x.
\end{equation}
Then, $T$ satisfies
\begin{equation}
(\divrg T)_x(\omega^2,\ldots,\omega^s) 
 = \int\limits_{P_x^+} (\pounds_L f)(x,p)\omega^2(p)\cdot\cdot\cdot\omega^s(p)\pi_x,
\qquad
\omega^2,\ldots,\omega^s\in T_x^*M.
\label{Eq:DivT}
\end{equation}
In particular, $T$ is divergence-free if $f$ satisfies the Liouville equation. In order to prove this, fix $s-1$ one-forms $\omega^2,\ldots,\omega^s$ on $M$ and replace the function $f(x,p)$ with the function $F(x,p) := f(x,p)\omega^2_x(p)\cdot\cdot\cdot\omega^s_x(p)$ on both sides of Eq.~(\ref{Eq:DivJBis}), noticing that this procedure replaces $J$ with the vector field $T(\cdot,\omega^2,\ldots,\omega^s)$ on $M$. Then, Eq.~(\ref{Eq:DivT}) follows by observing that $L[ \omega(p) ] = p^\mu p^\nu\nabla_\mu\omega_\nu$ in adapted local coordinates.

A particular important case is $s=2$, where $T$ represents the stress-energy tensor of the kinetic gas. Its most important properties are reviewed in Refs.~\cite{jE71,oStZ13}. In particular, it satisfies all the standard energy conditions as long as $f$ is nonnegative. Since it is symmetric and divergence-free (provided the Liouville equation holds), it allows one to describe a self-gravitating kinetic simple gas in $n\geq 3$ spacetime dimensions by considering the coupled Einstein-Liouville system of equations:
\begin{eqnarray}
G_{\mu\nu} &=& \kappa T_{\mu\nu} 
 = \kappa\int\limits_{P_x^+} f(x,p) p_\mu p_\nu\pi_x,
 \label{Eq:EinsteinLiouville1}\\
\pounds_L f &=& p^\mu\frac{\partial f}{\partial x^\mu}
 - \Gamma^\mu{}_{\alpha\beta}(x) p^\alpha p^\beta \frac{\partial f}{\partial p^\mu}
 = 0,
\label{Eq:EinsteinLiouville2}
\end{eqnarray}
with $G$ the Einstein tensor and $\kappa$ a coupling constant (equal $8\pi$ times Newton's constant when $n=4$). For recent work on properties and solutions of this system, see Refs.~\cite{hA11,Ringstrom-Book}.

\subsection{Newtonian limit}

Finally, we briefly discuss the Newtonian limit of Eqs.~(\ref{Eq:EinsteinLiouville1},\ref{Eq:EinsteinLiouville2}) in the four-dimensional case $n=4$. For this, we assume the existence of a local chart $(U,\phi)$ of the spacetime manifold $M$ satisfying the following conditions (see, for example, Ref.~\cite{Straumann-Book}):
\begin{enumerate}
\item[(i)] $g_{\mu\nu} = \eta_{\mu\nu} + h_{\mu\nu}$ with $(\eta_{\mu\nu}) = \diag(-1,+1,+1,+1)$ and $h_{\mu\nu}$ satisfying $|h_{\mu\nu}| \ll 1$
(weak gravitational field),
\item[(ii)] $|\partial_0 h_{\mu\nu}| \ll |\partial_i h_{\mu\nu}|$, $i=1,2,3$
(quasi-stationary gravitational field),
\item[(iii)] For each $x\in U$ the function $p\mapsto f(x,p)$ has its main support for those  $(p^\mu) = m\gamma (1,v^i)$ for which $|v^i| \ll 1$, $i=1,2,3$ (small velocities).
\end{enumerate}

Conditions (i) and (iii) imply that, to leading order in $h_{\mu\nu}$ and $v^i$, $\gamma\simeq 1$, which implies in turn that
\begin{displaymath}
\pi_x = \frac{\sqrt{-\det(g_{\mu\nu})}}{-\hat{p}_0} dp^1\wedge dp^2\wedge dp^3
\simeq \frac{1}{m} dp^1\wedge dp^2\wedge dp^3
\end{displaymath}
on the main support of $f(x,p)$. As a consequence, the energy density is approximately
\begin{displaymath}
T_{00}(x) \simeq m\int\limits_{\Real^3} f(x,p) d^3 p,
\end{displaymath}
while $T_{0j}$ and $T_{ij}$ are smaller by factors of $v^i$ and $v^i v^j$, respectively. Furthermore, using the three conditions (i),(ii), (iii) we find
\begin{displaymath}
\Gamma^\mu{}_{\alpha\beta} p^\alpha p^\beta \simeq -\frac{m^2}{2}\partial^i h_{00}
 = m^2\partial^i\Phi,
\end{displaymath}
where we have identified $-h_{00}/2 = \Phi$ with the Newtoninan potential $\Phi$. Consequently, in the Newtonian limit, the Einstein-Liouville system~(\ref{Eq:EinsteinLiouville1},\ref{Eq:EinsteinLiouville2}) reduces to the Poisson-Liouville system of equations:
\begin{eqnarray}
\Delta\Phi = 4\pi G_N\rho, && \qquad \rho(x) = m\int\limits_{\Real^3} f(x,p) d^3 p,
\label{Eq:PoissonLiouville1}\\
\frac{\partial f}{\partial t} + \frac{1}{m} p^i\frac{\partial f}{\partial x^i}
 + F^i\frac{\partial f}{\partial p^i} = 0, && \qquad F^i = -m\partial^i\Phi,
\label{Eq:PoissonLiouville2}
\end{eqnarray}
with $G_N$ Newton's constant. This system of equations plays an important role in stellar dynamics, see for example Ref.~\cite{BinneyTremaine-Book}.

\section{Symmetries of the distribution function}
\label{Sec:Symmetries}

In this section, we give a systematic discussion on the important issue of the symmetries of the distribution function. We start our discussion by considering a continuous symmetry $S$ of spacetime, that is, a one-parameter group of isometries of $(M,g)$ with associated Killing vector field $\xi$. We first show that $S$ can be lifted to $(TM,\hat{g})$. This defines a natural lift which maps $\xi$ to the infinitesimal generator $\hat{\xi}$ of a one-parameter group of isometries on $(TM,\hat{g})$. As it turns out, $\hat{\xi}$ is tangent to the mass shells $\Gamma_m$; therefore, its flow leaves each $\Gamma_m$ invariant. Once this generator $\hat{\xi}$ has been introduced, we define the distribution function $f$ to be $S$-symmetric by requiring $f$ to be invariant with respect to the flow generated by $\hat{\xi}$, that is $\pounds_{\hat{\xi}} f = 0$. Furthermore, we show that any generator $\hat{\xi}$ necessarily commutes with the Liouville vector field $L$, implying that the imposition of symmetries, together with the Liouville equation $\pounds_L f = 0$ does not yield additional restrictions.

Next, we consider a Lie group $G$ of isometries on the base manifold $(M,g)$. In this case, the symmetry group is generated by a finite number of Killing vector fields $\xi_1,\xi_2,\ldots,\xi_r$, satisfying commutation relations $[\xi_a,\xi_b] = C^d{}_{ab}\xi_d$ with associated structure constants $C^d{}_{ab}$. We show that the lift $\xi\mapsto \hat{\xi}$ preserves these commutation relations, that is, $[\hat{\xi}_a,\hat{\xi}_b] = C^d{}_{ab}\hat{\xi}_d$. Therefore, $G$ lifts naturally on $(TM,\hat{g})$. As in the previous case, we define the distribution function $f$ to be invariant under $G$ if $\pounds_{\hat{\xi}_a} f = 0$ for all $a=1,2,\ldots,r$.

Applications to spherically symmetric and stationary axisymmetric distribution functions will be discussed in the next section.

For related discussions about lifted transformations on the tangent bundle and symmetries, see Refs.~\cite{sS58,sS62} within the context of Riemannian geometry and~\cite{rMdT93,rMdT94} and references therein for applications to relativistic kinetic theory.

\subsection{One-parameter groups of symmetries}

Consider a one-parameter group of diffeomorphism $\varphi^\lambda: M \to M$ on the base manifold $M$ with associated vector field $\xi\in {\cal X}(M)$,
\begin{equation}
\xi_x := \left. \frac{d}{d\lambda} \right|_{\lambda=0} \varphi^\lambda(x),\qquad x\in M.
\end{equation}
We note that $\varphi^\lambda$ can be lifted naturally to $TM$, by defining
\begin{equation}
\hat{\varphi}^\lambda: TM\to TM,\quad (x,p)\mapsto
\hat{\varphi}^\lambda(x,p) := \left( \varphi^\lambda(x), \varphi^\lambda_{*x}(p) \right),
\end{equation}
where $\varphi^\lambda_{*x} : T_x M\to T_{\varphi^\lambda(x)} M$ denotes the push-forward of $\varphi^\lambda$ at $x$. It can easily be verified that $\hat{\varphi}^\lambda$ satisfies the following properties:
\begin{enumerate}
\item[(i)] $\pi\circ\hat{\varphi}^\lambda(x,p) = \varphi^\lambda(x)$,
\item[(ii)] $\hat{\varphi}^0(x,p) = (x,p)$
\item[(iii)] $\hat{\varphi}^\lambda\circ\hat{\varphi}^\mu(x,p) 
 = \hat{\varphi}^{\lambda+\mu}(x,p)$
\end{enumerate}
for all $(x,p)\in TM$, and thus it defines a one-parameter group of diffeomorphisms on $TM$. The associated infinitesimal generator is the vector field $\hat{\xi}\in {\cal X}(TM)$ defined as
\begin{equation}
\hat{\xi}_{(x,p)} 
 := \left. \frac{d}{d\lambda} \right|_{\lambda=0} \hat{\varphi}^\lambda(x,p),
\qquad (x,p)\in TM.
\label{Eq:xihat}
\end{equation}
In terms of adapted local coordinates $(x^\mu,p^\mu)$, the vector field $\hat{\xi}$ can be written as
\begin{equation}
\hat{\xi}_{(x,p)} = \xi^\mu(x) \left. \frac{\partial}{\partial x^\mu} \right|_{(x,p)} 
 + p^\alpha\frac{\partial \xi^\mu}{\partial x^\alpha}(x)
\left. \frac{\partial}{\partial p^\mu} \right|_{(x,p)},\qquad
\xi_x = \xi^\mu(x) \left. \frac{\partial}{\partial x^\mu} \right|_x.
\label{Eq:xihatLocCoord}
\end{equation}
In order to prove this, let $f\in {\cal F}(TM)$ be an arbitrary smooth function on $TM$, and let $(x,p)\in TM$. Then,
\begin{eqnarray*}
\hat{\xi}_{(x,p)}[f] 
 &=& \left. \frac{d}{d\lambda}\right|_{\lambda=0} f\left( \hat{\varphi}^\lambda(x,p) \right)
 \\
 &=& \left. \frac{d}{d\lambda}\right|_{\lambda=0} f\left( \varphi^\lambda(x), 
 \varphi_{*x}^\lambda(p) \right)\\
 &=& \xi^\mu(x)\left. \frac{\partial f}{\partial x^\mu} \right|_{(x,p)}
  + \frac{\partial \xi^\mu}{\partial x^\nu}(x) p^\nu 
  \left. \frac{\partial f}{\partial p^\mu} \right|_{(x,p)},
\end{eqnarray*}
where in the last step we have used the fact that if $y^\mu(\lambda,x)$ parametrizes the diffeomorphism $\varphi^\lambda(x)$, then the vector field $\varphi^\lambda_{*x}(p)$ has the components
\begin{displaymath}
\frac{\partial y^\mu}{\partial x^\nu}(\lambda,x) p^\nu,
\end{displaymath}
and consequently,
\begin{displaymath}
\left. \frac{d}{d\lambda}\right|_{\lambda=0} 
\frac{\partial y^\mu}{\partial x^\nu}(\lambda,x) p^\nu
 = \frac{\partial \xi^\mu}{\partial x^\nu}(x) p^\nu.
\end{displaymath}
The next result gives a covariant representation of $\hat{\xi}$.

\begin{lemma}
The vector field $\hat{\xi}$ defined in Eq.~(\ref{Eq:xihat}) can be written as
\begin{equation}
\hat{\xi} = (I^H)^{-1}(\xi) + (I^V)^{-1}(\nabla_p\xi).
\end{equation}
\end{lemma}

\proof By rewriting the term $\frac{\partial \xi^\mu}{\partial x^\alpha}$ in Eq.~(\ref{Eq:xihatLocCoord}) in terms of covariant derivatives, and by using the definition~(\ref{Eq:emu}) of $e_\mu$ it follows that
\begin{equation}
\hat{\xi}_{(x,p)} = \xi^\mu(x) \left. e_\mu \right|_{(x,p)} 
 + \nabla_p\xi^\mu(x)\left. \frac{\partial}{\partial p^\mu} \right|_{(x,p)},
\label{Eq:xihatLocCoordBis}
\end{equation}
from which the lemma follows.
\qed

Next, we show the following identity:

\begin{lemma}
\label{Lem:dHxi}
$dH_{(x,p)}(\hat{\xi}) = g_x(p,\nabla_p\xi)$ for all $(x,p)\in TM$.
\end{lemma}

\proof Using the local expressions~(\ref{Eq:dH},\ref{Eq:xihatLocCoordBis}) we find immediately
\begin{displaymath}
dH(\hat{\xi}) = g_{\mu\nu} p^\mu \theta^\nu(\hat{\xi}) 
 = g_{\mu\nu} p^\mu\nabla_p\xi^\nu = g(p,\nabla_p\xi).
\end{displaymath}
Alternatively, this important formula can also be derived without the use of local coordinates by appealing to the Hamiltonian property~(\ref{Eq:LHam}) of the Liouville vector field $L$,
\begin{eqnarray*}
dH(\hat{\xi}) &=& \Omega_s(\hat{\xi},L)\\
&=& \hat{g}(\hat{\xi},J(L))\\
&=& \hat{g}(\hat{\xi},(I^V)^{-1}(p))\\
&=& g(\nabla_p\xi,p),
\end{eqnarray*}
where we have used the definition~(\ref{Eq:Omegas}) of the symplectic form $\Omega_s$ in the second step, and the definitions of $J$, $L$ and $\hat{g}$ in the remaining steps. 
\qed

So far, the Killing property of the generator $\xi$ on $(M,g)$ has not been used. Therefore, the results so far hold for any one-parameter family of diffeomorphism. However, for the particular case for which $\varphi^\lambda$ is a one-parameter family of isometries of $(M,g)$, it follows from the previous lemma and the Killing property of $\xi$ that $dH(\hat{\xi}) = 0$, implying that $\hat{\xi}$ is tangent to the mass shells $\Gamma_m$. In particular, the lifted flow $\hat{\varphi}^\lambda$ leaves the mass shells invariant. 

We finish this subsection by noting that the lifted flow $\hat{\varphi}^\lambda$ on $(TM,\hat{g})$ is isometric if and only if the one-parameter group $\varphi^\lambda$ on $(M,g)$ is isometric.

\begin{proposition}[cf. \cite{sS58}]
Let $\xi\in {\cal X}(M)$ be a vector field on $M$ with corresponding lifted vector field $\hat{\xi}$ on $TM$. Then, $\xi$ is a Killing vector field on $(M,g)$ if and only if $\hat{\xi}$ is a Killing vector field on $(TM,\hat{g})$.
\end{proposition}

\proof We compute the components of $\pounds_{\hat{\xi}}\hat{g}$ with respect to the local basis of vector fields $\{ e_\mu, \frac{\partial}{\partial p^\mu} \}$ defined in Eq.~(\ref{Eq:emu}). For this, we evaluate the right-hand side of the identity
\begin{displaymath}
\pounds_{\hat{\xi}}\hat{g}(X,Y) = \hat{\xi}[ \hat{g}(X,Y) ] 
 - \hat{g}( \pounds_{\hat{\xi}} X, Y)
 - \hat{g}( X,\pounds_{\hat{\xi}} Y)
\end{displaymath}
for the particular vector fields $X,Y = e_\mu,\frac{\partial}{\partial p^\mu}$. Using the commutation relations
\begin{equation}
[e_\mu,e_\nu] = -R^\alpha{}_{\beta\mu\nu} p^\beta\frac{\partial}{\partial p^\alpha},
\quad
\left[ e_\mu, \frac{\partial}{\partial p^\nu} \right] = \Gamma^\alpha{}_{\mu\nu}
\frac{\partial}{\partial p^\alpha},
\quad
\left[ \frac{\partial}{\partial p^\mu} , \frac{\partial}{\partial p^\nu} \right] = 0
\end{equation}
and the representation of $\hat{\xi}$ in Eq.~(\ref{Eq:xihatLocCoordBis}) we find
\begin{eqnarray}
\pounds_{\hat{\xi}} e_\mu &=& [\hat{\xi},e_\mu]
 = -\frac{\partial\xi^\alpha}{\partial x^\mu} e_\alpha
 - p^\beta\left( \nabla_\mu\nabla_\beta\xi^\alpha 
   - R^\alpha{}_{\beta\mu\nu}\xi^\nu \right)
 \frac{\partial}{\partial p^\alpha},
\label{Eq:Lieemu}\\
\pounds_{\hat{\xi}} \frac{\partial}{\partial p^\mu} 
 &=& \left[\hat{\xi},\frac{\partial}{\partial p^\mu} \right]
 = -\frac{\partial\xi^\alpha}{\partial x^\mu} \frac{\partial}{\partial p^\alpha}.
\label{Eq:Liepmu}
\end{eqnarray}
Taking into account that $\hat{g}(e_\mu,e_\nu) = \hat{g}\left( \frac{\partial}{\partial p^\mu},\frac{\partial}{\partial p^\nu} \right) = g_{\mu\nu}$, $\hat{g}\left(e_\mu,\frac{\partial}{\partial p^\nu}\right) = 0$, we obtain from this
\begin{eqnarray*}
&& \pounds_{\hat{\xi}}\hat{g}(e_\mu,e_\nu)
 = \pounds_{\hat{\xi}}\hat{g}\left( \frac{\partial}{\partial p^\mu},\frac{\partial}{\partial p^\nu} \right) 
 = \pounds_\xi g_{\mu\nu},\\
&& \pounds_{\hat{\xi}}\hat{g}\left( e_\mu,\frac{\partial}{\partial p^\nu} \right) 
 = p^\beta\left( \nabla_\mu\nabla_\beta\xi_\nu - R_{\nu\beta\mu\alpha}\xi^\alpha \right).
\end{eqnarray*}
Therefore, if $\xi$ is a Killing vector field of $(M,g)$, then $\pounds_\xi g = 0$ and $\nabla_\mu\nabla_\beta\xi_\nu - R_{\nu\beta\mu\alpha}\xi^\alpha = 0$ which implies that $\pounds_{\hat{\xi}}\hat{g} = 0$. Conversely, if $\hat{\xi}$ is a Killing vector field of $(TM,\hat{g})$ then $\pounds_{\hat{\xi}}\hat{g} = 0$ which implies $\pounds_\xi g = 0$.
\qed

\subsection{Symmetries of the distribution function}

Let $\xi\in {\cal X}(M)$ be a Killing vector field on $(M,g)$ generating a continuous symmetry $S$ of spacetime, that is, a one-parameter family of isometries $\varphi^\lambda: M \to M$. We define the distribution function $f$ to be $S$-symmetric if it is invariant with respect to the lifted flow $\hat{\varphi}^\lambda: TM\to TM$. Equivalently, this means that
\begin{equation}
\pounds_{\hat{\xi}} f = 0.
\label{Eq:fSym}
\end{equation}
It is worth mentioning that in principle, one could have considered the lift of an arbitrary one-parameter family of diffeomorphism $\varphi^\lambda$ on $M$, regardless of whether or not it is an isometry group of $(M,g)$. However, this prescription is generally accompanied by additional strong restrictions. For example, if $f$ is required to satisfy the Liouville equation $\pounds_L f = 0$, then one obtains the constraint
\begin{equation}
\pounds_{[L,\hat{\xi}]} f = 0.
\label{Eq:LSymConstraint}
\end{equation}
A direct calculation based on adapted local coordinates and the commutation relation~(\ref{Eq:Lieemu}) reveals that
\begin{equation}
[L,\hat{\xi}] = p^\alpha p^\beta\left[Ê\nabla_\alpha\nabla_\beta\xi^\mu
 - R^\mu{}_{\beta\alpha\nu}\xi^\nu \right] \frac{\partial}{\partial p^\mu}.
\label{Eq:LSymCommutator}
\end{equation}
If $\xi$ is a Killing vector field of $(M,g)$, then the expression inside the square parenthesis vanishes, and the constraint~(\ref{Eq:LSymConstraint}) is automatically satisfied. Therefore, if $\xi$ generates a continuous symmetry $S$ of the spacetime manifold $(M,g)$, then the requirement of $f$ to be $S$-symmetric is compatible with the Liouville equation and does not lead to further restrictions.

Connected with these observation we close this subsection with the following general remarks regarding the symmetries of the symplectic structure.

\begin{proposition}
\label{Prop:Canonical}
Let $\xi\in {\cal X}(M)$ be the generator of a one-parameter group of isometries on $(M,g)$. Then, the lifted vector field $\hat{\xi}$ satisfies the identity
\begin{equation}
i_{\hat{\xi}}\Omega_s = -dP,\qquad
P := \Theta(\hat{\xi}) = g(p,\xi),
\end{equation}
where $\Theta$ is the Poincar\'e one-form defined in Eq.~(\ref{Eq:PoincareOneForm}). In particular, $\hat{\xi}$ is the infinitesimal generator of a symplectic flow on $TM$, that is $\pounds_{\hat{\xi}}\Omega_s = 0$.
\end{proposition}

\proof One possible way of proving the statement is to utilize the coordinate expressions~(\ref{Eq:OmegasLocCoord},\ref{Eq:xihatLocCoordBis}) and $P = g_{\mu\nu} p^\mu\xi^\nu$. A straightforward calculation then yields
\begin{displaymath}
i_{\hat{\xi}}\Omega_s + dP = p^\mu(\nabla_\mu\xi_\nu + \nabla_\nu\xi_\mu) dx^\nu,
\end{displaymath}
and the right-hand side vanishes since $\xi$ is a Killing vector field on $(M,g)$. Applying the exterior derivative operator $d$ on both sides of this equation and using the fact that $d\Omega_s = 0$ yields $\pounds_{\hat{\xi}}\Omega_s = 0$, which shows that $\hat{\xi}$ generates a symplectic flow on $TM$.

An alternative way of proving the result is to use $\Omega_s = d\Theta$ and Cartan's identity to find
\begin{displaymath}
i_{\hat{\xi}}\Omega_s = \pounds_{\hat{\xi}}\Theta - dP.
\end{displaymath}
A short calculation reveals that $\pounds_{\hat{\xi}}\Theta_{(x,p)}(X) = \pounds_\xi g_x (p,\pi_{*(x,p)}(X)) = 0$ for all $(x,p)\in TM$ and all $X\in T_{(x,p)}(TM)$. 
\qed

An important consequence of Proposition~\ref{Prop:Canonical} is that the quantity $P_{(x,p)} = g_x(p,\xi)$ is conserved along the Liouville flow if $\xi$ is a Killing vector field of $(M,g)$:

\begin{proposition}
\label{Prop:Conservation}
Let $\xi\in {\cal X}(M)$ be the generator of a one-parameter group of isometries on $(M,g)$. Then, the quantity $P_{(x,p)} := g_x(p,\xi)$ is conserved along the flow generated by the Liouville vector field.
\end{proposition}

\proof Using Proposition~\ref{Prop:Canonical}, Lemma~\ref{Lem:dHxi} and the fact that $\xi$ is a Killing vector field it follows that
\begin{displaymath}
\pounds_L P = dP(L) = -\Omega_s(\hat{\xi},L) = \Omega_s(L,\hat{\xi}) = -dH(\hat{\xi}) = 0.
\end{displaymath}
\qed

A different way of stating this result is to say that the Poisson bracket between $H$ and $P$ is zero, $\{ H,P \} = 0$. Using standard arguments from Hamiltonian mechanics, this implies that the corresponding Hamiltonian vector fields $L$ and $\hat{\xi}$ commute with each other, $[L,\hat{\xi}] = 0$, offering an alternative explanation for the compatibility of the condition~(\ref{Eq:fSym}) with the Liouville equation.

\subsection{Lie groups of symmetries}

After discussing one-parameter groups of isometries, we now consider Lie groups $G$ of isometries of the spacetime $(M,g)$. These are generated by a finite number of Killing vector fields $\xi_1,\xi_2,\ldots,\xi_r$ satisfying the commutation relations
\begin{equation}
[\xi_a,\xi_b] = C^d{}_{ab}\xi_d,\qquad
a,b = 1,2,\ldots r,
\end{equation}
with $C^d{}_{ab}$ the structure constants associated to $G$. The next result implies that the lifted Killing vector fields $\hat{\xi}_1,\hat{\xi}_2,\ldots,\hat{\xi}_r$ satisfy exactly the same commutation relations, implying that $G$ also acts as an isometry group on $(TM,\hat{g})$.

\begin{lemma}
\label{Lem:Comm}
Let $\xi,\eta\in {\cal X}(M)$. Then,
\begin{equation}
[\hat{\xi},\hat{\eta}] = \hat{\zeta},\qquad
\zeta = [\xi,\eta].
\end{equation}
\end{lemma}

\proof This can be verified in a straightforward way by making use of the local coordinate expression~(\ref{Eq:xihatLocCoord}) for $\hat{\xi}$.
\qed

By analogy to the case of a $S$-symmetric distribution function $f$,
a $G$-symmetric distribution function is defined by $\pounds_{\hat {\xi}_a} f = 0$ for $a=1,2,\ldots,r$. These conditions are compatible with the Liouville equation $\pounds_L f = 0$ in view of the comments following Eq.~(\ref{Eq:LSymCommutator}).

We shall make use of these properties in the next section, when discussing spherically symmetric and stationary, axisymmetric distribution functions.

\section{Applications}
\label{Sec:Applications}

In this section, we apply the theory developed so far to the discussion of two particular scenarios in four-dimensional general relativity. In the first case we consider spherically symmetric distribution functions on spherically symmetric spacetimes. Such systems are of relevance in astrophysical situations such as the description of globular clusters, the distribution of stars around non-rotating supermassive black holes and various models of collisionless spherical accretion. In the second case we derive the most general collisionless distribution function on a Kerr black hole background, including the particular case where this distribution function is stationary and axisymmetric. Such distributions are highly relevant for the modeling of astrophysical processes near an isolated rotating black hole or the description of distributions of stars around supermassive black holes at the center of galaxies.

\subsection{Spherically symmetric distribution functions}

We consider a spherically symmetric spacetime manifold $(M,g)$. This has the form $M = \tilde{M}\times S^2$ with metric
\begin{equation}
g = \tilde{g}_{ab} dx^a\otimes dx^b + r^2\gamma_{AB} dx^A\otimes dx^B,
\label{Eq:gSphSym}
\end{equation}
where $(\tilde{M},\tilde{g} := \tilde{g}_{ab} dx^a\otimes dx^b)$ is a two-dimensional pseudo-Riemannian manifold, $(S^2,\gamma:=\gamma_{AB} dx^A\otimes dx^B)$ is the unit metric sphere and $r\in {\cal F}(\tilde{M})$ is a strictly positive and smooth function on $\tilde{M}$ representing the areal radius of the invariant two-spheres. It is defined geometrically as
\begin{displaymath}
r(x) = \sqrt{\frac{A(x)}{4\pi}},
\end{displaymath}
where $A(x)$ is the area of the sphere $\{ x \} \times S^2$ through $x\in \tilde{M}$. Here and in the following, the coordinates $x^a$, $a=0,1$, are local coordinates on $\tilde{M}$, while $x^A$, $A=2,3$, are local coordinates on $S^2$.

By construction, the metric~(\ref{Eq:gSphSym}) is $SO(3)$-symmetric. The corresponding generators can be written, in terms of spherical coordinates $(x^A) = (\vartheta,\varphi)$, as
\begin{eqnarray*}
\xi_1 &:=& -\sin\varphi\frac{\partial}{\partial \vartheta} 
 - \cot\vartheta\cos\varphi\frac{\partial}{\partial \varphi},\\
\xi_2 &:=& \cos\varphi\frac{\partial}{\partial \vartheta} 
 - \cot\vartheta\sin\varphi\frac{\partial}{\partial \varphi},\\
\xi_3 &:=& \frac{\partial}{\partial \varphi},
\end{eqnarray*}
and they satisfy the commutation relations
\begin{displaymath}
[\xi_1,\xi_2] = -\xi_3,\qquad\hbox{and cyclic permutations of $123$}.
\end{displaymath}
In terms of adapted local coordinates $(x^a,\vartheta,\varphi,p^a,p^\vartheta,p^\varphi)$ the corresponding lifted generators are given by
\begin{eqnarray*}
\hat{\xi}_1 &:=& -\sin\varphi\frac{\partial}{\partial \vartheta} 
 - \cot\vartheta\cos\varphi\frac{\partial}{\partial \varphi}\\
&-& \cos\varphi\; p^\varphi\frac{\partial}{\partial p^\vartheta}
 + \left( \frac{\cos\varphi}{\sin^2\vartheta} p^\vartheta 
 + \cot\vartheta\sin\varphi\; p^\varphi \right)\frac{\partial}{\partial p^\varphi},\\
\hat{\xi}_2 &:=& \cos\varphi\frac{\partial}{\partial \vartheta} 
 - \cot\vartheta\sin\varphi\frac{\partial}{\partial \varphi}\\
&-& \sin\varphi\; p^\varphi\frac{\partial}{\partial p^\vartheta}
 + \left( \frac{\sin\varphi}{\sin^2\vartheta} p^\vartheta 
 - \cot\vartheta\cos\varphi\; p^\varphi \right)\frac{\partial}{\partial p^\varphi},\\
\hat{\xi}_3 &:=& \frac{\partial}{\partial \varphi},
\end{eqnarray*}
where in deriving these expressions we have used the expression~(\ref{Eq:xihatLocCoord}). According to Lemma~\ref{Lem:Comm} these generators satisfy the same commutation relations as the $\xi_a$'s, that is
\begin{displaymath}
[\hat{\xi}_1,\hat{\xi}_2] = -\hat{\xi}_3,\qquad\hbox{and cyclic permutations of $123$},
\end{displaymath}
as can also be verified by a lengthy explicit calculation.

According to our definition in the previous section, a $SO(3)$-symmetric distribution function $f$ obeys the conditions
\begin{equation}
\pounds_{\hat{\xi}_a} f = 0,\qquad a=1,2,3.
\end{equation}
The condition $\pounds_{\hat{\xi}_3} f = 0$ implies that $f$ is independent of $\varphi$ while $\pounds_{\hat{\xi}_1 + i\hat{\xi}_2} f = 0$ is equivalent to
\begin{displaymath}
e^{i\varphi}\left[ i\frac{\partial f}{\partial \vartheta} 
 - \cot\vartheta\frac{\partial f}{\partial \varphi}
 - p^\varphi\frac{\partial f}{\partial p^\vartheta}
 + \left( \frac{p^\vartheta}{\sin^2\vartheta} - i\cot\vartheta\, p^\varphi \right)
 \frac{\partial f}{\partial p^\varphi} \right] = 0,
\end{displaymath}
which yields the two conditions
\begin{displaymath}
\frac{\partial f}{\partial \vartheta} 
 = \cot\vartheta\, p^\varphi\frac{\partial f}{\partial p^\varphi},\qquad
p^\varphi\frac{\partial f}{\partial p^\vartheta} 
 = \frac{p^\vartheta}{\sin^2\vartheta}\frac{\partial f}{\partial p^\varphi}.
\end{displaymath}
This implies that $f$ must have the following form:
\begin{equation}
f(x,p) = F(x^a,p^a,\ell),\quad
\ell := r^2\sqrt{\gamma_{AB} p^A p^B} 
 = r^2\sqrt{(p^\vartheta)^2 + \sin^2\vartheta (p^\varphi)^2}.
\label{Eq:SphSymf}
\end{equation}
Here, the quantity $\ell$ has a direct physical interpretation: since the vector fields $\xi_a$, $a=1,2,3$, are Killing vector fields of $(M,g)$ it follows from Proposition~\ref{Prop:Conservation} that the three quantities
\begin{eqnarray*}
\ell_1(x,p) &:=& g_x(p,\xi_1) = r^2\left( -\sin\varphi\, p^\vartheta 
 - \cos\vartheta\sin\vartheta\cos\varphi\, p^\varphi \right),\\
\ell_2(x,p) &:=& g_x(p,\xi_2) = r^2\left( \cos\varphi\, p^\vartheta 
 - \cos\vartheta\sin\vartheta\sin\varphi\, p^\varphi \right),\\
\ell_3(x,p) &:=& g_x(p,\xi_3) = r^2\sin^2\vartheta\, p^\varphi,
\end{eqnarray*}
are invariant with respect to the Liouville flow. The quantity $\ell$ defined in Eq.~(\ref{Eq:SphSymf}) is the total angular momentum $\ell = \sqrt{\ell_1^2 + \ell_2^2 + \ell_3^2}$  which is also invariant under the Liouville flow.

Next, let us evaluate the Liouville operator on a spherically symmetric distribution function. According to Eq.~(\ref{Eq:SphSymf}) we have
\begin{eqnarray}
L[f] &=& \frac{\partial F}{\partial x^a} L[x^a] + \frac{\partial F}{\partial p^a} L[p^a]
 + \frac{\partial F}{\partial \ell} L[\ell]
\nonumber\\
 &=& p^a\frac{\partial F}{\partial x^a} - \left( \tilde{\Gamma}^a{}_{cd} p^c p^d
 - \frac{\ell^2}{r^2}\frac{r^a}{r} \right) \frac{\partial F}{\partial p^a},
\label{Eq:LiouvilleSphSym}
\end{eqnarray}
where we have used the fact that $L[\ell]=0$ and the expressions
\begin{displaymath}
\Gamma^a{}_{cd} = \tilde{\Gamma}^a{}_{cd},\qquad
\Gamma^a{}_{cD} = 0,\qquad
\Gamma^a{}_{CD} = -r r^a\gamma_{CD}
\end{displaymath}
for the Christoffel symbols, the tilde referring to the two-metric $\tilde{g}_{ab}$ and $r^a$ to $\tilde{g}^{ab}\tilde{\nabla}_b r$. Also, we remind the reader that the indices $a,b,c,d,\ldots$ run over $0,1$ while the Capital indices $C,D$ refer to angular coordinates. The reduction in Eq.~(\ref{Eq:LiouvilleSphSym}) suggests that in spherical symmetric the theory can be formulated as an effective theory on the tangent bundle $T\tilde{M}$ of the two-dimensional manifold $(\tilde{M},\tilde{g})$. There are two possible interpretations for this effective theory. In the first one, one considers $\ell$ as a parameter and works with the family of distribution functions $F_\ell(x^a,p^a) = F(x^a,p^a,\ell)$ on the effective mass shell determined by the condition
\begin{equation}
\tilde{g}_{ab} p^a p^b = -m^2 - \frac{\ell^2}{r^2}.
\end{equation}
In the second interpretation one trades off the centrifugal term $\ell^2/r^2$ in Eq.~(\ref{Eq:LiouvilleSphSym}) by the term $-m^2 - \tilde{g}_{ab} p^a p^b$ and works on the closed submanifold $\tilde{g}_{ab} p^a p^b \leq -m^2$ of the tangent bundle $T\tilde{M}$.

\subsection{Current density and stress-energy tensor for a spherical distribution function}

Next, we evaluate the current density $J$ and stress-energy tensor $T$ for a spherically symmetric distribution function, i.e. a distribution function of the form $f(x,p) = F(x^a,p^a,\ell)$ where $x^a$, $p^a$, $\ell $ have been defined in the previous subsection. To evaluate $J$ and $T$ it is convenient to introduce an orthonormal basis of vectors $\{ e_0,e_1,e_2,e_3 \}$ on the tangent space $T_x M$. Relative to such a basis the components of $J$ and $T$ have the form
\begin{eqnarray*}
J^\alpha(x) = \int\limits_{P_x^+} f(x,p) p^\alpha \pi_x,
&& \hbox{(current density)},\\
T^{\alpha\beta}(x) = \int\limits_{P_x^+} f(x,p) p^\alpha p^\beta \pi_x,
&& \hbox{(stress-energy tensor)},
\end{eqnarray*}
where here the orthonormal coordinates $(p^\alpha) = (p^0,p^j)$ on $T_x M$ are defined by $p = p^\alpha e_\alpha$ for $p\in T_x M$, and the invariant volume element $\pi_x$ has the special-relativistic Lorentz-invariant form
\begin{displaymath}
\pi_x = \frac{dp^1\wedge dp^2\wedge dp^3}{\sqrt{m^2 + \delta_{ij} p^i p^j}}.
\end{displaymath}
For the spherically symmetric metric~(\ref{Eq:gSphSym}) this volume element of the future mass hyperboloid $P_x^+$ can be rewritten in an equivalent form that is adapted to the $2+2$-structure $\tilde{M}\times S^2$ of the background spacetime. For this, we first orient the orthonormal frame such that $(p^0,p^1)$ and $(p^2,p^3)$ are adapted to this structure. For example, in terms of spherical coordinates on $S^2$, we choose $(p^0,p^1)$ to be orthonormal coordinates on $T_x\tilde{M}$ and $p^2 = r p^\vartheta$, $p^3 = r\sin\vartheta p^\varphi$. Noting that $r\sqrt{(p^2)^2 + (p^3)^2} = \ell$ is the total angular momentum, we reexpress $(rp^2,rp^3)$ in terms of polar coordinates $(\ell,\phi)$ such that
\begin{displaymath}
dp^{2}\wedge dp^{3} =\frac{\ell}{r^2} d\ell \wedge d\phi.
\end{displaymath}

As we have mentioned at the end of the last subsection, the reduction of the Liouville operator in Eq.~(\ref{Eq:LiouvilleSphSym}) suggests that in spherical symmetry we can either work with a family of distribution functions of the form $F_\ell(x^a,p^a)$ or equivalently trade off the centrifugal term $\ell^2/r^2$ in Eq.~(\ref{Eq:LiouvilleSphSym}) by the term $-m^2 - \tilde{g}_{ab} p^a p^b$ and work on the closed submanifold $\tilde{g}_{ab} p^a p^b \leq -m^2$ of the tangent bundle $T\tilde{M}$. Here, we chose to work
with the second interpetation and thus we eliminate  the $\ell$-contribution from the volume form $\pi_x$. Differentiating the relation
\begin{displaymath}
\frac{\ell^2}{r^2} = -m^2 + (p^0)^2 - (p^1)^2
\end{displaymath}
yields
\begin{displaymath}
\frac{\ell d\ell}{r^2} = p^0 dp^0 - p^1 dp^1.
\end{displaymath}
Combining the above formulas together and noticing that on the future mass hyperboloid $p^0 = \sqrt{m^2 + \delta_{ij} p^i p^j}$ we obtain simply
\begin{displaymath}
\pi_x = \frac{dp^1\wedge dp^2 \wedge dp^3}{\sqrt{m^2 + \delta_{ij} p^i p^j}}
 = -dp^0\wedge dp^1\wedge d\phi.
\end{displaymath}
Using this and taking $f(x,p) = F(x^a,p^0,p^1)$ we find the following components of the current density in terms of the orthonormal frame $\{ e_0,e_1,e_2,e_3 \}$,
\begin{subequations}
\begin{eqnarray}
J^0 &=& -2\pi\int\limits_{{\hat P}_x^+}  F(x^a,p^0,p^1)p^0 dp^0\wedge dp^1,\\
J^1 &=& -2\pi\int\limits_{{\hat P}_x^+}  F(x^a,p^0,p^1)p^1 dp^0\wedge dp^1,\\
J^A &=& 0,\qquad A=2,3.
\end{eqnarray}
\end{subequations}
In a similar manner we find that the non-vanishing components of the stress-energy tensor are as follows\footnote{The minus sign in front of the integrals reflects the fact that the coordinate system $(p^0,p^1,\phi)$ has the opposite orientation than the original system $(p^1,p^2,p^3)$.}
\begin{subequations}
\begin{eqnarray}
T^{00} &=& 
 -2\pi\int\limits_{{\hat P}_x^+}  F(x^a,p^0,p^1)(p^0)^2 dp^0\wedge dp^1,\\
T^{01} =T^{10} &=& 
 -2\pi\int\limits_{{\hat P}_x^+}  F(x^a,p^0,p^1)p^0p^1 dp^0\wedge dp^1,\\
T^{11} &=&
 -2\pi\int\limits_{{\hat P}_x^+}  F(x^a,p^0,p^1)(p^1)^2 dp^0\wedge dp^1,\\
T^{22} = T^{33} &=& 
 -\pi\int\limits_{{\hat P}_x^+}  F(x^a,p^0,p^1)\left[ (p^0)^2 - (p^1)^2 - m^2 \right] dp^0\wedge dp^1,
\end{eqnarray}
\end{subequations}
where the integration should be performed in the region ${\hat P_{x}}$ of the $(p^0, p^1)$-plane defined by $(p^0)^2 - (p^1)^2 \geq m^2$ and $p^0 > 0$. These conditions suggest that it might be advantageous to parametrize $(p^0,p^1)$ in terms of a radial-like coordinate $P$ subject to $P\geq m$ and a hyperbolic angle $\chi\in\Real$ such that $(p^0,p^1) = P(\cosh\chi,\sinh\chi)$. In terms of this the non-vanishing orthonormal components of $J$ and $T$ are
\begin{subequations}
\begin{eqnarray}
J^0 &=& 2\pi\int\limits_m^\infty dP P^2\int\limits_{-\infty}^\infty d\chi\cosh\chi
  F(x^a,p^0,p^1),\\
J^1 &=& 2\pi\int\limits_m^\infty dP P^2\int\limits_{-\infty}^\infty d\chi\sinh\chi
 F(x^a,p^0,p^1),
\end{eqnarray}
\end{subequations}
and
\begin{subequations}
\begin{eqnarray}
T^{00} &=& 2\pi\int\limits_m^\infty dP P^3\int\limits_{-\infty}^\infty d\chi\cosh^2\chi
  F(x^a,p^0,p^1),\\
T^{01} = T^{10} 
 &=& 2\pi\int\limits_m^\infty dP P^3\int\limits_{-\infty}^\infty d\chi\cosh\chi\sinh\chi
  F(x^a,p^0,p^1),\\
T^{11} &=& 2\pi\int\limits_m^\infty dP P^3\int\limits_{-\infty}^\infty d\chi\sinh^2\chi
  F(x^a,p^0,p^1),\\
T^{22} = T^{33}
 &=& \pi\int\limits_m^\infty dP P(P^2 - m^2)\int\limits_{-\infty}^\infty d\chi F(x^a,p^0,p^1).
\end{eqnarray}
\end{subequations}

The expressions for the stress-energy tensor can be combined with the corresponding expressions for the Einstein tensor for a spherically symmetric metric of the form~(\ref{Eq:gSphSym}), see for instance Ref.~\cite{eCnOoS13}. Together with the reduction~(\ref{Eq:LiouvilleSphSym}) of the Liouville operator, this yields the following system of equations:
\begin{eqnarray}
&& \frac{1}{r}\tilde{\nabla}^a\tilde{\nabla}^b r - \frac{m_{MS}}{r^3}\tilde{g}^{ab}
 = -4\pi G_N\left[ T^{ab} + \tilde{g}^{ab}(T^{00} - T^{11}) \right],
\quad a,b = 0,1,\qquad\\
&& \frac{\tilde{\Delta} r}{r} - \tilde{k} = 4\pi G_N(T^{22} + T^{33}),
\\
&& p^a\frac{\partial F}{\partial x^a} 
- \left[ \left( \tilde{\Gamma}^a{}_{cd} + \frac{r^a}{r}\tilde{g}_{cd} \right)p^c p^d
+ m^2\frac{r^a}{r} \right]\frac{\partial F}{\partial p^a} = 0,
\end{eqnarray}
where here $\tilde{\Delta} r := \tilde{g}^{ab}\tilde{\nabla}_a\tilde{\nabla}_b r$, $\tilde{k}$ is the Gauss curvature of $(\tilde{M},\tilde{g})$ and $m_{MS}$ is the Misner-Sharp mass function~\cite{cMdS64} which is defined invariantly by $\tilde{g}^{ab} r_a r_b = 1 - 2m_{MS}/r$.

\subsection{Most general collisionless distribution function on a Kerr black hole}
\label{Sec:Kerr}

As a second application of our formalism, we derive the most general collisionless distribution function on a Kerr black hole background. Our goal is to solve the Liouville equation $\pounds_L f = 0$ on a Kerr background $(M,g)$ with mass parameter $m_H$ and rotation parameter $a_H$. We restrict ourself to the black hole case $a_H^2 \leq m_H^2$. The strategy is to find a symplectic transformation on $TM$ which trivializes the Hamiltonian and the Liouville vector field. This can be achieved by finding a complete solution of the Hamilton-Jacobi equation
\begin{equation}
H(x,p) = -\frac{1}{2} m^2,\qquad p = \nabla S,
\end{equation}
where $S\in {\cal F}(M)$ is the generating function on $(M,g)$ and $\nabla S$ its gradient. Using the definition~(\ref{Eq:HDefBis}) of the Hamiltonian function, we can also write this equation explicitly as
\begin{equation}
g_x(\nabla S,\nabla S) = g^{-1}_x(dS,dS) = -m^2.
\label{Eq:HJ}
\end{equation}
For the case of the Kerr spacetime, the Hamilton-Jacobi equation is known to be separable~\cite{bC68} due to the Killing structure of spacetime~\cite{mWrP70}. 
In terms of Boyer-Lindquist coordinates~\cite{MTW-Book} Eq.~(\ref{Eq:HJ}) reads
\begin{eqnarray}
&-& \frac{1}{\Delta} \left[
 (r^2+a_H^2)\frac{\partial S}{\partial t} + a_H\frac{\partial S}{\partial\varphi} \right]^2
 + \left[ \frac{1}{\sin\vartheta}\frac{\partial S}{\partial\varphi}
  + a_H\sin\vartheta\frac{\partial S}{\partial t} \right]^2
\nonumber\\
&+& \Delta\left( \frac{\partial S}{\partial r} \right)^2 
 + \left( \frac{\partial S}{\partial\vartheta} \right)^2 
 = -m^2\left(  r^2 + a_H^2\cos^2\vartheta \right),
\label{Eq:KerrHJ}
\end{eqnarray}
where the quantity $\Delta$ is defined as
\begin{displaymath}
\Delta := r^2 - 2m_H r + a_H^2.
\end{displaymath}
The Kerr metric has the two commutating Killing vector fields
\begin{displaymath}
k = \frac{\partial}{\partial t},\qquad
l = \frac{\partial}{\partial\varphi}
\end{displaymath}
generating isometries with respect to time translations and rotations, respectively. As a consequence of Proposition~\ref{Prop:Conservation}, the quantities
\begin{equation}
E := -p_t = -g(k,p),\qquad
\ell_z := p_\varphi = g(l,p)
\end{equation}
are constant along the trajectories. Furthermore, it can be seen from Eq.~(\ref{Eq:KerrHJ}) that the problem is separable in the variables $r$ and $\vartheta$~\cite{bC68,MTW-Book}. Therefore, we make the following ansatz for the generating function:
\begin{equation}
S = -E t + \ell_z\varphi + S_r(r) + S_\vartheta(\vartheta).
\end{equation}
Substituting this ansatz into Eq.~(\ref{Eq:KerrHJ}) yields
\begin{eqnarray*}
&& \frac{1}{\Delta} \left[ (r^2+a_H^2)E - a_H\ell_z \right]^2 - m^2 r^2 - \Delta S_r'(r)^2
\\
&=& \left( \frac{\ell_z}{\sin\vartheta} - a_H\sin\vartheta E \right)^2 
 + m^2 a_H^2\cos^2\vartheta
 + S_\vartheta'(\vartheta)^2.
\end{eqnarray*}
Since the left-hand side depends only on $r$ and the right-hand side only on $\vartheta$ it follows that
\begin{eqnarray*}
S_r'(r)^2 &=& \frac{R(r)}{\Delta(r)^2},\quad
R(r) := \left[ (r^2+a_H^2)E - a_H\ell_z \right]^2 - \Delta(m^2 r^2 + \ell^2),\\
S_\vartheta'(\vartheta)^2 &=& \Theta(\vartheta),\quad
\Theta(\vartheta) := \ell^2 - \left( \frac{\ell_z}{\sin\vartheta} - a_H\sin\vartheta E \right)^2
  - m^2 a_H^2\cos^2\vartheta,
\end{eqnarray*}
where $\ell^2$ is the Carter constant.\footnote{Note that in the non-rotating limit, $\ell^2 = p_\vartheta^2 + p_\varphi^2/\sin^2\vartheta$, so $\ell$ is the total angular momentum in this case.} Therefore, the generating function is, formally,
\begin{equation}
S(t,\varphi,r,\vartheta,m,E,\ell_z,\ell) = -E t + \ell_z\varphi
 + \int\limits^r \sqrt{R(r)} \frac{dr}{\Delta(r)}
 + \int\limits^\vartheta \sqrt{\Theta(\vartheta)} d\vartheta.
\end{equation}
According to the general theory of Hamilton-Jacobi~\cite{Arnold-Book}, this function generates a symplectic transformation
\begin{equation}
(t,\varphi,r,\vartheta,p_t,p_\varphi,p_r,p_\vartheta) \mapsto (Q^0,Q^1,Q^2,Q^3,P_0,P_1,P_2,P_3),
\label{Eq:SymplecticTransformation}
\end{equation}
where the new variables $(Q^a,P_a)$ are defined as
\begin{subequations}
\begin{eqnarray}
&& P_0 := m,\qquad P_1 := E,\qquad P_2 := \ell_z,\qquad P_3 := \ell,\\
&& Q^0 := \frac{\partial S}{\partial m} 
 = -m\int\limits^r\frac{r^2 dr}{\sqrt{R(r)}} 
 - m a_H^2\int\limits^\vartheta\frac{\cos^2\vartheta d\vartheta}{\sqrt{\Theta(\vartheta)}},\\
&& Q^1 := \frac{\partial S}{\partial E} 
 = -t + \int\limits^r\frac{ (r^2 + a_H^2)A(r)}{\sqrt{R(r)}}\frac{dr}{\Delta(r)}
 + a_H\int\limits^\vartheta\frac{B(\vartheta)}{\sqrt{\Theta(\vartheta)}}
 d\vartheta,\\
&& Q^2 := \frac{\partial S}{\partial \ell_z} 
= \varphi - a_H\int\limits^r\frac{A(r)}{\sqrt{R(r)}}\frac{dr}{\Delta(r)}
 - \int\limits^\vartheta\frac{B(\vartheta)}{\sqrt{\Theta(\vartheta)}}
 \frac{d\vartheta}{\sin^2\vartheta},\\
&& Q^3 := \frac{\partial S}{\partial \ell} 
 = -\ell\int\limits^r\frac{dr}{\sqrt{R(r)} }
 + \ell\int\limits^\vartheta\frac{d\vartheta}{\sqrt{\Theta(\vartheta)}},
\end{eqnarray}
\end{subequations}
with the functions $A(r) := (r^2 + a_H^2)E - a_H\ell_z$ and $B(\vartheta) := \ell_z - a_H\sin^2\vartheta E$. In fact, it can be easily verified that the transformation~(\ref{Eq:SymplecticTransformation}) leaves the symplectic form in Eq.~(\ref{Eq:OmegasLocCoord}) invariant:
\begin{displaymath}
\Omega_s = dp_\mu\wedge dx^\mu = dP_\alpha\wedge dQ^\alpha.
\end{displaymath}
Since $H = -m^2/2 = -P_0^2/2$, it follows from the Hamiltonian equations that the quantities $Q^1,Q^2,Q^3,P_0,P_1,P_2,P_3$ are constant along the trajectories and that
\begin{equation}
\dot{Q}^0 = \frac{\partial H}{\partial m} = -m.
\end{equation}
Consequently, the Liouville vector field in these new coordinates assumes the simple form
\begin{equation}
L = -m\frac{\partial}{\partial Q^0}.
\end{equation}
This offers an enormous simplification for the determination of a collisionless distribution function. In terms of the new coordinates the Liouville equation is equivalent to the statement that the distribution function is an arbitrary function independent of $Q^0$:
\begin{equation}
f(x,p) = F(Q^1,Q^2,Q^3,P_0,P_1,P_2,P_3).
\label{Eq:fKerr}
\end{equation}

Notice also that in these new coordinates the natural lift defined in Eq.~(\ref{Eq:xihat}) of the Killing vector fields $k$ and $l$ is simply
\begin{equation}
\hat{k} = -\frac{\partial}{\partial Q^1},\qquad
\hat{l} = \frac{\partial}{\partial Q^2}.
\end{equation}
Therefore, a collisionless distribution function required to be stationary and axisymmetric has the same form as in Eq.~(\ref{Eq:fKerr}), with the function $F$ being independent of $Q^1$ and $Q^2$.

It should be mentioned that our analysis in arriving at Eq.~(\ref{Eq:fKerr}) has been formal. Indeed, the generating function $S$, and as a consequence also the functions $Q^0,\ldots Q^3$ are multi-valued in general. Therefore, appropriate periodicity conditions on $F$ need to be specified in order for the distribution function to be well-defined. Specific applications will be discussed in future work. For a generalization to a collisionless charged gas propagating on a Kerr-Newman black hole background, see Ref.~\cite{oStZ14a}.

\section{Conclusions}
\label{Sec:Conclusions}

In this work, utilizing the rich geometrical structure that the tangent bundle $TM$ acquires from the spacetime $(M,g)$, we presented a geometrically oriented approach to relativistic kinetic theory. Due to the natural metric $\hat{g}$ on $TM$, a free of ambiguities, natural integration theory has been developed which in turn offers new insights in the interpretation of the theory. Specifically, a relativistic gas from the perspective of the tangent bundle can be interpreted as the flow of an incompressible fictitious fluid in the mass shell with the distribution function playing the role of a particle density. Together with the Liouville vector field, the distribution function defines a current density on the mass shell which, for a collisionless gas, is conserved. This current gives rise to a physical current density on the spacetime manifold which is also conserved and represents the first moment of the distribution function. Although this interpretation of the distribution function is different from the one arising from the Hamiltonian framework discussed in~\cite{oStZ13}, the two approaches are identical and complement each other.

Furthermore, the bundle metric $\hat g$ offers the means to provide a useful connection between the symmetries of the background spacetime and the symmetries of the distribution function. Specifically, we have shown that groups $G$ of isometries of $(M,g)$ lift naturally to groups of isometries of $(TM,\hat g)$. In addition, we have shown that these lifted isometries give rise to symplectic flows on the tangent bundle. A distribution function is defined to be $G$-symmetric if it is invariant with respect to the lifted isometries. As an application of this definition, we have derived the most general spherically symmetric distribution function on an arbitrary spherical spacetime. In this case, the Liouville equation reduces to an effective equation on the tangent bundle of the two-dimensional radial manifold $\tilde{M}$. We have also reduced the Einstein-Liouville system to an effective problem on this manifold.

As a second application, based on the Hamiltonian structure of the theory, we have constructed the most general collisionless distribution function on a Kerr black hole spacetime. This construction arises by taking into account the spacetime symmetries of the Kerr metric combined with the separability of the Hamilton-Jacobi equation. This approach offers the opportunity of modeling several astrophysically interesting scenarios, including distributions of stars around a supermassive black hole and accretion of a dilute gas into a rotating black hole.

Recently~\cite{oStZ14a}, we have extended our framework to the case of a simple, collisionless charged gas. In this work, the distribution function satisfies the Einstein-Maxwell-Vlasov system of equations. In the Newtonian limit, this system reduces to the Poisson-Maxwell-Vlasov system. For a recent interesting application of this system to the modeling of current loops and the generated magnetic fields, see Ref.~\cite{cCzS13}.

In this work we have restricted our analysis to a collisionless simple gas, i.e. a collection of classical particles of the same rest mass $m > 0$ moving on future directed timelike geodesics of the background spacetime. In future work, we are aiming to extend the analysis to collisional systems of dilute gases described by the relativistic Boltzmann equation. In the light of the analysis of the present paper, the kinetic theory of a gas described by the Boltzmann equation raises many interesting questions. For instance, an important issue is the relation between the symmetries of the background spacetime and the distribution function with those of the collision integral. Moreover, a system which is interesting from an astrophysical and cosmological point of view consists of a photon gas. As we have already mentioned, a tangent bundle formulation of this  system raises some delicate questions due the fact that the Liouville vector field has critical points, a property that is in sharp contrast with the case of a simple gas of particles with positive rest mass.

\acknowledgments
We thank Pierre Bayard for fruitful discussions regarding the geometry of the tangent bundle. This work was supported in part by CONACyT Grant No. 101353 and by a CIC Grant to Universidad Michoacana.

\appendix
\section{Proof of Proposition~\ref{Prop:LGeodesic}}
 
In this appendix we prove that the Liouville vector field $L$ satisfies the geodesic equation $\hat{\nabla}_L L = 0$ on the tangent bundle without explicitly computing the Levi-Civita connection of the bundle metric $\hat{g}$. To this purpose we first show that the Poincar\'e one-form defined in Eq.~(\ref{Eq:PoincareOneForm}) is dual to the Liouville vector field.

\begin{lemma}
\label{Lem:Theta}
$\Theta = \hat{g}(L,\cdot)$.
\end{lemma}

\proof Since $L$ is horizontal, we have for each $(x,p)\in TM$ and $X\in T_{(x,p)}(TM)$,
\begin{displaymath}
\hat{g}_{(x,p)}(L_{(x,p)},X) = g_x(\pi_{*(x,p)}(L_{(x,p)}),\pi_{*(x,p)}(X)) 
= g_x(p,\pi_{*(x,p)}(X)) = \Theta_{(x,p)}(X).
\end{displaymath}
\qed

From this, we also obtain the following interesting identity:
\begin{lemma}
\label{Lem:dH}
$\pounds_L\Theta = dH$.
\end{lemma}

\proof Using the Cartan identity, the result from the previous lemma and $\Omega_s = d\Theta$ we find
\begin{displaymath}
\pounds_L\Theta = di_L\Theta + i_L d\Theta = d[ \hat{g}(L,L) ] + i_L\Omega_s
 = 2dH - dH = dH,
\end{displaymath}
where we have also used the fact that $L$ is the Hamiltonian vector field associated to $H$, see Eq.~(\ref{Eq:LHam}).
\qed

Based on these results, we can easily prove Proposition~\ref{Prop:LGeodesic}. For this, let $X$ be any vector field on $TM$. Then, using Lemma~\ref{Lem:Theta} and the Ricci identity we first find
\begin{eqnarray*}
(\pounds_L\Theta)(X) &=& L[ \Theta(X) ] - \Theta(\pounds_L X)\\
 &=& L[\hat{g}(L,X)] - \hat{g}(L,\pounds_L X)\\
 &=& \hat{g}(\hat{\nabla}_L L,X) + \hat{g}(L,\hat{\nabla}_L X - [L,X]).
\end{eqnarray*}
Next, using the fact that $\hat{\nabla}$ is torsion-free and the definition~(\ref{Eq:HDef}) of $H$,
\begin{displaymath}
\hat{g}(L,\hat{\nabla}_L X - [L,X]) = \hat{g}(L,\hat{\nabla}_X L)
 = \frac{1}{2} X[ \hat{g}(L,L) ] = dH(X).
\end{displaymath}
Combining these results with the identity in Lemma~\ref{Lem:dH} we conclude that $\hat{g}(\hat{\nabla}_L L, X) = 0$ for all vector fields $X$ on $TM$, and Proposition~\ref{Prop:LGeodesic} follows.

\bibliographystyle{unsrt}
\bibliography{refs_kinetic}

\end{document}